\documentclass[acmsmall,screen]{acmart}

\AtBeginDocument{%
  }

\setcopyright{acmlicensed}
\copyrightyear{2025}
\acmYear{2025}
\acmDOI{XXXXXXX.XXXXXXX}

\acmConference[Conference CSCW '25]{The 2025 ACM SIGCHI Conference on Computer-Supported Cooperative Work \& Social Computing}{October 18 -- 22, 2025}{Bergen, Norway}
\acmISBN{978-1-4503-XXXX-X/18/06}

\usepackage{booktabs}
\usepackage{xargs} 
\usepackage{soul}  
\usepackage{color} 
\usepackage{xspace}
\usepackage{array}
\usepackage{graphicx}
\usepackage{pdfpages}
\usepackage{listings}
\usepackage{enumitem}
\usepackage{verbatim}
\usepackage{makecell}
\newcommand{\eg}{{\it e.g.,\ }}

\newcommand{\ie}{{\it i.e.,\ }}

\usepackage{booktabs}
\usepackage{array}
\usepackage{multirow}
\definecolor{oxfordblue}{rgb}{0.0, 0.13, 0.28}
\definecolor{harvardcrimson}{rgb}{0.79, 0.0, 0.09}
\definecolor{dartmouthgreen}{rgb}{0.05, 0.5, 0.06}
\definecolor{princetonorange}{rgb}{1.0, 0.56, 0.0}
\definecolor{yaleblue}{rgb}{0.06, 0.3, 0.57}
\definecolor{usccardinal}{rgb}{0.6, 0.0, 0.0}
\definecolor{uclablue}{rgb}{0.33, 0.41, 0.58}
\definecolor{msugreen}{rgb}{0.09, 0.27, 0.23}
\definecolor{cornellred}{rgb}{0.7, 0.11, 0.11}
\definecolor{pomegranate}{RGB}{192, 57, 43}
\definecolor{anti-pomegranate}{RGB}{43,178,192}
\definecolor{alizarin}{RGB}{231, 76, 60}
\definecolor{anti-belize}{RGB}{185, 41, 56}
\definecolor{belize}{RGB}{41, 128, 185}
\definecolor{peter}{RGB}{52, 152, 219}
\definecolor{green}{RGB}{22, 160, 133}
\definecolor{anti-green}{RGB}{160,22,118}
\definecolor{turquoise}{RGB}{26, 188, 156}
\definecolor{pumpkin}{RGB}{211, 84, 0}
\definecolor{anti-pumpkin}{RGB}{0,22,211}
\definecolor{carrot}{RGB}{230, 126, 34}
\definecolor{wisteria}{RGB}{142, 68, 173}
\definecolor{anti-wisteria}{RGB}{99,173,68}
\definecolor{amethyst}{RGB}{155, 89, 182}
\definecolor{nephritis}{RGB}{39, 174, 96}
\definecolor{anti-nephritis}{RGB}{174,39,117}

\newcommand{\zhenhui}[1]{{\color{black} #1}}

\newcommand{\peng}[1]{{\color{black} #1}}
\newcommand{\name}{{\textit{ComViewer}}} 
\newcommand{\shiwei}[1]{{\color{black} #1}}
\newcommand{\vis}{{\textit{Zoomable Posts}}}
\newcommand{\note}{{\textit{Note-taking}}}
\newcommand{\question}{\textit{Questioning}}

\newcommand{\chuhan}[1]{{\color{black} #1}}
\newcommand{\mingxiang}[2]{{\color{black} #1}}
\newcommand{\revision}[1]{{\color{black} #1}}
\newcommand{\proofreading}[1]{{\color{black} #1}}
\begin{document}

\title{ComViewer: An Interactive Visual Tool to Help Viewers Seek Social Support in Online Mental Health Communities}

\author{Shiwei Wu}
\email{wushw28@mail2.sysu.edu.cn}
\affiliation{%
  \institution{Sun Yat-sen University}
  \city{Zhuhai}
  \country{China}
}
\author{Mingxiang Wang}
\email{wangmingxiang@hnu.edu.cn}
\affiliation{%
  \institution{Hunan University}
  \city{Changsha}
  \country{China}
}
\author{Chuhan Shi}
\email{chuhanshi@seu.edu.cn}
\affiliation{%
  \institution{Southeast University}
  \city{Nanjing}
  \country{China}
}
\author{Zhenhui Peng}
\authornote{Corresponding author.}
\email{pengzhh29@mail.sysu.edu.cn}
\affiliation{%
  \institution{Sun Yat-sen University}
  \city{Zhuhai}
  \country{China}
}



\begin{abstract}
Online mental health communities (OMHCs) offer rich posts and comments for viewers, who do not directly participate in the communications, to seek social support from others' experience. 
However, viewers could face challenges in finding helpful posts and comments and digesting the content to get needed support, as revealed in our formative study (N=10). 
In this work, we present an interactive visual tool named \name{} to help viewers seek social support in OMHCs. 
With \name{}, viewers can filter posts of different topics and find supportive comments via a zoomable circle packing visual component that adapts to searched keywords. 
Powered by LLM, \name{} supports an interactive sensemaking process by enabling viewers to interactively highlight, summarize, and question any community content. 
A within-subjects study (N=20) demonstrates \name{}'s strengths in providing viewers with a more 
\revision{simplified}, more fruitful, and more engaging support-seeking experience compared to a baseline OMHC interface without \name{}. 
We further discuss design implications for facilitating information-seeking and sense making in online mental health communities. 
\end{abstract}



\received{20 February 2007}
\received[revised]{12 March 2009}
\received[accepted]{5 June 2009}

\maketitle

\section{Introduction}
Online mental health communities (OMHCs) offer rich user-generated content, \zhenhui{\eg posts that describe health conditions and comments that provide social support}, for viewers to learn from others. 
In OMHCs, without the necessity to create a post or leave a comment, viewers can find people who have experienced similar mentally challenging issues and gather information to handle these mental health issues \cite{schildkraut2021survivor}. 
By browsing the posts and comments in OMHCs, viewers can get emotional support (\ie empathy, love, or concern) and informational support (\ie information or advice) \cite{sharma2018mental} to 
\revision{cope with their mental health concerns.}

However, seeking social support from the posts and comments in OMHCs could be challenging for many viewers.
For one thing, 
the posts in OMHCs vary across different topics (\eg school, job, sleepless \cite{kim2023supporters}) and purposes (\eg seeking informational or emotional support \cite{10.1145/3411764.3445446}), and the comments are of varying quality \cite{agichtein2008finding,xu2012you,xu2014voyant,marlow2014rookie,peng2020exploring}. 
OMHCs commonly enable users to specify queried keywords, choose a tap (\eg flairs in Reddit \cite{redditcom}), and select sorting criteria (\eg hot, new) to filter needed content. 
Nevertheless, the filtered results based on these features may not be able to match the topics and offer social support that users actually need, especially when they do not have a clear mind at the beginning or find it difficult to articulate their needs in text because they lack domain-specific language \cite{palani2021conotate}. 
Besides, traditional OMHCs sites present the original posts and subsequent replies as a paginated indented list which only supports a linear way of navigation \cite{hoque2014convis}.
Thus viewers need to go through a long list of comments sequentially until their information needs are fulfilled \cite{hoque2014convis, peltonen2017topic}, 
which is time-consuming for information seeking \cite{liu2019unakite, liu2021reuse}.
For another, 
\zhenhui{upon finding interested posts and comments, viewers lack support from the OMHCs to actively mark the useful content and summarize the takeaways, which help viewers take actions to alleviate their mental health concerns \cite{de2017language}.} 
Besides, many comments in OMHCs could have offered useful suggestions but may lack necessary background information, rationales, or actionable plans for viewers to \zhenhui{make sense} of the suggestions \cite{omari2016novelty,fujita2020diverse}.

\zhenhui{Researchers in human-computer interaction have explored the usage of visualization techniques \cite{tu2022phrasemap, hoque2014convis,7192683, wu2014opinionflow} to support information seeking in online communities and large language models (LLMs) for improving sense making \cite{suh2023sensecape, jiang2023graphologue}. 
For example, 
\shiwei{
VisOHC \cite{kwon2015visohc} visualizes individual online health community (OHC) conversation threads for OHC administrators to understand the quality of individual postings,  thereby facilitating monitoring of the communities.
}
However, little work attempts to provide visual information-seeking assistance for viewers with mental health concerns, who could have different information-seeking needs and challenges compared to the target users (\eg community moderators \cite{wu2014opinionflow}, medicinal chemists \cite{shi2022medchemlens}) of previous visual analytic systems. 
As for sense making, previous tools like Sensecape \cite{suh2023sensecape} and Graphlogue \cite{jiang2023graphologue} have supported users to make sense of the LLM-generated content via multilevel abstraction and graphical diagrams. 
Nevertheless, the design and impact of sense-making support with LLMs on user-generated content in online communities are still under-explored. 
}

\zhenhui{
In this paper, we propose \name{}, an interactive visual tool that facilitates viewers to seek social support in OMHCs by helping them navigate needed posts and make sense of the helpful content in comments to 
\revision{cope with their mental health concerns.}}
We first conducted a formative study in which ten viewers with recent mental health concerns explored r/Anxiety \footnote{\url{https://www.reddit.com/r/Anxiety/}}, a representative OMHC in Reddit, to seek \zhenhui{social support about their anxious issues.}  
\shiwei{
The findings revealed viewers' demand to visually search posts and comments based on various dimensions (\eg topics and the amounts of contained social support) as well as interactive organization and questioning with the domain-specific content for sensemaking. 
}
Based on the derived design requirements, we designed three main panels in \name{} to satisfy viewers' demands. 
To start with, viewers can input any keywords to search related posts, which will be visually presented in the \vis{} panel (\autoref{fig:all} A). 
In this panel, they can zoom in and zoom out the circles and click the support filters (\autoref{fig:all} A3) to explore the searched posts of different topics, sought social support \proofreading{of the posts}, 
and provided social support in the comments under the posts.
\zhenhui{When viewers select a post to read it in detail, they can actively highlight any content of their interests, organize and summarize it with an LLM in the \note{} panel (\autoref{fig:all} B), and question it with an LLM in the \question{} panel (\autoref{fig:all} C).} 

We conducted a within-subject study with twenty participants to evaluate \name{}'s effectiveness and user experience. 
The results showed that compared to the baseline condition in which participants browsing an OMHC without \zhenhui{our proposed three panels in \name{}}, participants in the \name{} condition gained significantly more informational support and were significantly more engaging within the support-seeking process. 
Participants deemed \name{} significantly more useful than the baseline, highlighting the benefits of its \vis{} panel for 
\revision{simplifying the process of}
finding desired posts, its \note{} panel for managing learned knowledge, and its \question{} panel for making sense of the comments.
We further discuss design considerations for supporting information-seeking with visualization techniques and supporting information management with LLMs in online mental health communities. 

Our contributions to CSCW communities are three folds. 
First, we propose an interactive visual tool named \name{} that aids viewers in navigating and making sense of posts and comments to seek needed social support in online mental health communities. 
Second, our within-subjects study with 20 participants demonstrates the effectiveness of \name{} in improving support-seeking outcomes and experience compared to a baseline community interface without our proposed three panels in \name{}. 
Third, we offer insights into facilitating information seeking and \zhenhui{sense making} in online communities with visualization techniques and large language models. 

\textbf{Ethics and Researcher Disclosure}. 
We shape the work by our experience with and observation on people who struggle with stressful issues and seek social support online. 
The authors have experience of seeking and providing social support in online mental health communities. 
\shiwei{One author who has published work on OMHCs was trained by the university to cope with students' mental health issues.} 
Our research team obtains IRB approval for broader research projects on patients' and caregivers' practices of healthcare service systems and online communities.
We do not include any personally identifiable information such as username, gender, and age in the collected OMHC dataset. 
Besides, we secure the data in firewalled servers, and researchers could download the data only on local machines. 
\shiwei{In the design process and evaluation of \name{}, we monitor each participant's mental state, inform them that they can quit the study at any time if they feel uncomfortable about it, and suggest them to seek professional healthcare services if they face severe problems. 
We do not target and include participants who are officially diagnosed with mental health diseases but recommend them to seek professional help rather than using \name{}}. 
\section{Related Work}
\subsection{Social Support in Online Mental Health Communities}

Online Mental Health Communities (OMHCs) provide a convenient platform for people with mental health challenges to exchange social support with peers \cite{10.1145/2818048.2819994, info:doi/10.2196/jmir.1933}.
For example, in Reddit r/Anxiety, an anonymous OMHC with over \zhenhui{691K members up to June 2024, one can create a support-seeking post about encountering stressful issues, and others can respond by text under the post to offer their help \cite{peng2020exploring}}.
Support derived from these communities has been found to casually improve mental well-being like reduced likelihood of suicidal thoughts \cite{de2017language}.
Studies have shown that OMHCs play an important role in providing support seekers with emotional support (\eg empathy, acknowledgment) and informational support (\eg advice, suggestion) \cite{sharma2018mental}.
They can not only benefit members who create posts or leave comments \cite{saha2020causal, 10.1145/3290605.3300294} but also help viewers to explore posts with similar mental health problems \cite{huh2016personas}.
Broadly speaking, OMHCs have an abundance of information and members' experiences that viewers can learn from to alleviate their mental health issues, as illustrated by the paraphrased quote above.


\zhenhui{There are two main stages in the viewers' support-seeking process in OMHCs. First, they need to find posts and comments that contain desired information. Second, they need to make sense of and make use of content in the posts to alleviate their mental health problems. 
Both stages can be challenging, as the user-generated content in online communities, or general social media platforms, is highly heterogeneity, with huge volume, and fasting changing rate \cite{wu2016survey,liu2019unakite, liu2021reuse}.
The posts and comments are commonly unstructured and of various quality, \eg regarding the types of feedback \cite{peng2023designquizzer,peng2020exploring}, which can hinder viewers from quickly locating and getting useful information. 
All these challenges can increase the mental effort required for viewers who have mental health issues during the support-seeking process. 
What's more, viewers' needs may vary according to their own experiences and feelings \cite{huh2016personas}, while the user-generated content in OMHCs alone may not be sufficient to fulfill individual information needs \cite{omari2016novelty,fujita2020diverse}. 
}

\zhenhui{
Our work is motivated by the benefits of OMHCs for viewers to seek needed social support and aims to assist them in this support-seeking process. 
}

\zhenhui{
\subsection{Visual Information-Seeking Support and its Usage in Online Communities}
The considerable noisy data in online communities can easily overwhelm people and prevent them from acquiring meaningful information \cite{wu2016survey}. 
Traditionally, online communities present the retrieved information in a ranked list of documents \cite{hoque2014convis}. Users need to scan through the list and pick relevant documents from the top of the ranking, which could be a mentally demanding information-seeking process. 
A promising solution to ease this process is to visualize the search result space \cite{chen2000empirical}. 
In terms of supporting information seeking in online communities, researchers have proposed various related visual analytic tools.
For example, \citet{kwon2015visohc} 
\proofreading{built}
a visual \proofreading{analytic} tool named VisOHC to support administrators in understanding the dynamics of conversations that happen in online health communities (OHCs).
It represents individual conversation threads as visual collapsed boxes, which can assist OHC administrators in evaluating the quality of individual postings, such that they can learn how to improve their service in the communities.
\citet{hoque2014convis} 
\proofreading{proposed}
a visual exploratory text analytic system named ConVis, which visualizes clustered topics and mined opinions along with a set of metadata to aid readers in exploring and analyzing social media data.
Besides, another tool named OpinionFlow \cite{wu2014opinionflow} 
\proofreading{was developed to combine}
a Sankey graph with an improved density map to visually convey the flow of opinions among users. Such a visual feature can help domain experts trace and analyze the diffusion of opinions on social media in large-scale events \cite{wu2014opinionflow}.

Nevertheless, \chuhan{previous tools do not} attempt to address the information-seeking needs of viewers with mental health concerns, who may encounter challenges and look for information different from the target users of previous visual tools (\eg medicinal chemists \cite{shi2022medchemlens}, community moderators \cite{kwon2015visohc, wu2014opinionflow}).
Our work fills this gap by identifying the information-seeking challenges of viewers in OMHCs and designing and evaluating a visual \vis{} panel to address these challenges. 
}
\zhenhui{
\subsection{Information Sense Making with Large Language Models}
}
Large language models (LLMs) with their ability to instantly generate high-quality responses to flexible natural language queries may change the way people make sense of information \cite{suh2023sensecape}.
Millions of users are now turning to LLMs (\eg ChatGPT) for information-seeking tasks \cite{yen2024search, zhang2024see, tanprasert2024debate}, \zhenhui{such as finding explanations and summarizing content \cite{gero2024attribute, li2024chatcite}.}
\zhenhui{
Previous studies have explored support sense making of the responses generated by LLMs.
For example, Sensecape \cite{suh2023sensecape} 
enables
users to engage in information-seeking tasks in a more structured and systematic way compared to the traditional chat interface of LLMs.  
Specifically, it supports the flexible organization of the gathered information via multiple levels of abstraction in an information space for the LLM-generated content, and users can interactively explore the content in different levels by further prompting the LLM. 
Graphlogue \cite{jiang2023graphologue} converts text-based responses from LLMs into graphical diagrams to facilitate information-seeking, concept explanation and question-answering tasks.
It indicates that the node-link diagrams enhanced users' understanding of the diverse relationships inherent in the topic they explored.
Besides, \citet{sensemaking24Gero} 
\proofreading{presented}
many LLM responses at once which captures one or more aspects of LLM responses' possible variation and consistency to help make sense of LLM responses according to the \proofreading{users'} prompt.

All these previous works suggest that LLMs can be leveraged to aid viewers in making sense of information via interactive information exploration, organization, and comprehension.
However, little work looks at supporting sense making of the user-generated content in online communities with LLMs.
Compared to sense making of only the LLM-generated content, our scenario has unique challenges as it requires users to iterate reading and understanding unfamiliar content encountered in the community without expert guidance  \cite{Selenite24Liu}. 
Our work seeks to fill this gap by designing a
\proofreading{\note{} and \question{}}
panel in \name{} and examining its usefulness for facilitating sense making of the user-generated content in OMHCs with LLM-generated content. 



}

\section{Formative Study}
To understand viewers' challenges and needs for support in the process of seeking social support in online mental health communities (OMHCs), we \proofreading{conducted} a formative study with ten participants. 
The findings of this study inform the design choices we \proofreading{made} for \name{}. 

\zhenhui{
\subsection{Research Site} \label{sec:research_site}
\zhenhui{
We 
\proofreading{selected}
one of the mental health communities in Reddit, r/Anxiety, as the given OMHC for the participants. 
Reddit is an important, distinguished, convenient, and popular online community to share mental-health-related experiences and ``surprisingly high-quality'' information compared to other social media platforms like Facebook and Twitter (now X) \cite{de2014mental}.
Other OMHCs, such as 7cups, despite their emphasis more on one-on-one anonymous chat or professional therapy services, also contain similar peer-to-peer support content to that in OMHCs in Reddit.
The r/Anxiety is a public place of discussion and support for sufferers and loved ones with anxiety conditions. 
Despite the differences in focused mental health issues, r/Anxiety shares commonality with other Reddit OMHCs (e.g., r/depression) that members share informational and emotional support within the community \cite{sharma2018mental}. 
We will discuss the generality and limitations of our proposed tool and findings in the Discussion section (\autoref{sec:discussion}). 
}
}

\zhenhui{
\subsection{Participants and Procedure}
}
We recruited ten students (S01-S10, 5 females, 5 males, age range 20-25) from a local university, a representative group that often struggles with mental health \cite{park2018social}, via word-of-mouth. 
The inclusion criteria were that they indicated recent experiences with mentally challenging issues, had experience in viewing posts in OMHCs (\eg support groups in Weibo) to seek social support, and had competency in reading in English in the background survey. 
\zhenhui{
The test on Generalized Anxiety Disorder 7 (score range 0 - 21) indicated that on average ($Mean = 9.3, SD = 5.5$), our participants had mild anxiety ($score >= 5$).
Five participants exhibited moderate ($score >= 10$) or severe anxiety ($score >= 14$).
Four participants (S01, 02, 03, 05) reported feelings of anxiety related to the study, two (S04, 06) indicated concerns about their relationship, two (S07, 09) expressed worries about their physical health, one (S08) \proofreading{expressed} anxiety due to difficulty in determining whether to pursue a PhD, and one (S10) reported anxiety due to difficulties in finding a suitable job.
As for participants' experiences in the OMHCs, one participant reported that he used OMHCs daily, three used OMHCs 4-6 times a week, three used OMHCs at least once a week, and three seldom used OMHCs.
}

We conducted a one-on-one formative study with each participant. 
The study started with querying participants' mentally challenging issues in detail and introducing the task of browsing the posts and comments in r/Anxiety to seek social support. 
Then, we asked participants to explore r/Anxiety \zhenhui{for 30 minutes} and share their support outcomes with us.
Next, we conducted a semi-structured interview with them about their practices of seeking social support in the OMHC, the challenges encountered in this process, and the expected technical support for facilitating this process. 
Each interviewee spent around an hour in our study and received 60 RMB as compensation.
The interviews were recorded, transcribed, and analyzed using the reflexive thematic analysis method \cite{braun2012thematic} \zhenhui{by two authors.
}

\subsection{Findings} \label{sec: finding}
\subsubsection{Practices of Seeking Social Support by Browsing the OMHC}
In Reddit r/Anxiety, our participants actively searched posts and read comments about anxiety in research, job hunting, and social relationships. 
All participants agreed that the OMHC had useful content that provided them with emotional and informational support. 
All participants reported that finding others with similar experiences 
\proofreading{made}
them feel less isolated and realize that their situations are not unique. 
\revision{
Six participants also mentioned that reading about similar experiences or receiving relevant advice could help address their concerns, even if the posts did not exactly match their specific problems, as they thought the solutions to some mental health concerns were often similar.}
Additionally, seven participants appreciated the supportive comments they read, which offered suggestions and spiritual support that 
\revision{helped cope with their mentally challenging concerns.}
\zhenhui{
For example, S09 shared 7 suggestions from the community she thought could be operated on and tried (\eg ``\textit{The use of essential oils to regulate emotional states}'').} 

\begin{figure}[]
  \centering
  \includegraphics[width=\linewidth]{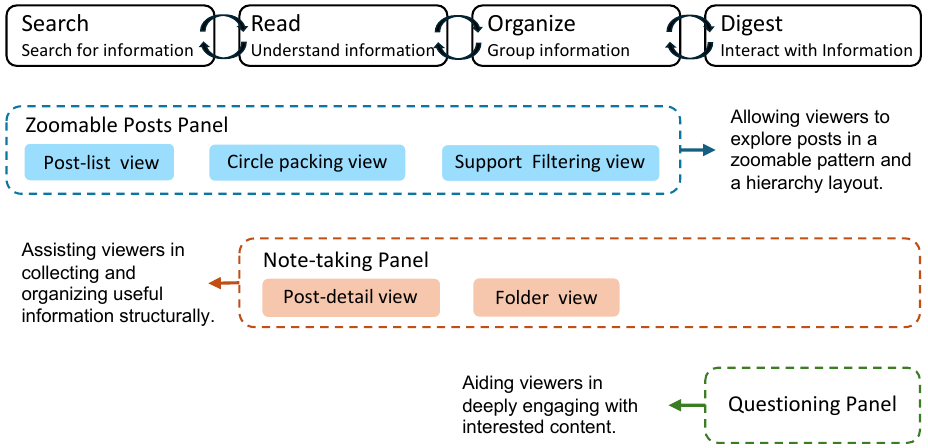}
  \caption{\zhenhui{A common information-seeking flow in online mental health communities. \name{} supports this flow with a Zoomable Posts Panel, a Note-taking Panel, and a Questioning Panel.}}
  \label{fig:pro_pic}
\end{figure}

\zhenhui{
As illustrated in \autoref{fig:pro_pic}, participants' information-seeking process commonly contains the following four steps: \textit{search}, \textit{read}, \textit{organize}, and \textit{digest}.
First, participants input keywords that describe their mental health issues in the search box to search for related information.
``\textit{I searched for keywords that I thought could describe my problem, like `friend', `hurt', and `say something bad', to see if the search results matched my needs}'' (S07, F, 23).
Then, they click on the returned posts whose titles catch their eyes, read the content of the post and its associated comments, and evaluate whether the content is relevant and helpful to their issues.
They would iterate the keywords in the search box if the read content is mostly irrelevant or unhelpful. 
``\textit{I read the posts and found they mostly talk about helping people with depression find jobs that do not match what I'm looking for. So I removed `depression' from my search terms}'' (S10, M, 25). 
Next, when encountering helpful content, 
participants may take notes and organize it for future usage.
``\textit{While reading comments, I would copy helpful comments into sticky notes and sort suggestions to consider how they could help alleviate anxiety about my research}'' (S09, F, 22).
After that, they digest their noted content to 
\revision{cope with their mental health concerns.}
``\textit{While digging into the content of my interest, I think of how to solve my problems accordingly}'' (S5, M, 22). 
Participants experience these four steps back and forth in their information-seeking process in the community.
}

\subsubsection{Challenges of Seeking Social Support by Browsing the OMHC}
\begin{table*}[]
    \centering
     \caption{Viewers' challenges in seeking social support from posts and comments in online mental health communities in the formative study.}
    \label{tab:formativeStudyResults}
    \begin{tabular}{p{2cm}<{\centering}p{3.5cm}p{5.9cm}p{1.2cm}<{\centering}}
    \toprule
     Theme & Category & Code example & Mentions \\ \midrule
     \multirow{9}{*}{\makecell[c]{Finding\\ desired\\ posts and\\ comments \\ (Search\&Read \\ steps)}} & {C1. Articulating mental health issues in keywords} & {``\textit{I'm not sure how to explain my mental health issues more clearly because I don't have the right words to do so.}'' (S03, F, 23)} & \multirow{3}{*}{8/10} \\
     \cline{2-4}
     ~ & {C2. Finding posts that are similar to the explored one} & {``\textit{I'd like to see if the comments under other similar posts have a better solution to solve my issues.}'' (S02, F, 22)}& \multirow{3}{*}{7/10} \\
     \cline{2-4}
     ~ & {C3. Filtering comments that provide the needed type of social support} & {``\textit{I'm looking for some more practical advice to help me solve the problem, rather than just spiritual comfort.}'' (S04, M, 23)} & \multirow{3}{*}{7/10} \\ \midrule
     \multirow{8}{*}{\makecell[c]{Making sense \\of the helpful \\content \\ (Organize\&\\Digest steps)}} & \multirow{4}{*}{\shortstack{C4. Organizing useful \\content to \revision{cope with} \\mental health \revision{concerns}}}  & {``\textit{The useful points I see are all small and fragmented, and it is impossible to form a relatively complete solution to the problem.}'' (S05, M, 22)} & \multirow{4}{*}{5/10} \\
     \cline{2-4}
     ~ & \multirow{4}{*}{\shortstack{C5.  Making sense of   \\ domain-specific content}} & {``\textit{I come across some mental health terms that I don’t understand, and I need to jump to other pages to find possible explanations.}'' (S09, F, 23)} & \multirow{4}{*}{6/10} 
    \\ \bottomrule
    \end{tabular}
\end{table*}

\zhenhui{In total, we \proofreading{coded} 96 pieces of data regarding encountered challenges into 5 categories and 2 themes, as shown in \autoref{tab:formativeStudyResults}.}

\textbf{Finding desired posts and comments in the Search\&Read steps 
\revision{takes considerable effort in refining search criteria}.
} 
\shiwei{
For all participants, to find desired posts and comments, they need to search and read posts under different search keywords, during which participants face three challenges.
First, eight participants reported difficulty in \textit{articulating their mental health issues in keywords} (\textbf{C1}) during the \textit{search} step.
Four participants said that they did not know what to search at the beginning, as they ``\textit{are new to the community and are unfamiliar with the topics discussed in the community}'' indicated by S02 (F, 23). 
After reading the returned posts in detail, eight participants were not satisfied with the first search results. So they had to think of other keywords and repeat the \textit{search} and \textit{read} step. 
S04 (M, 22) stated the difficulty in thinking of suitable keywords that describe his problem.
``\textit{People have different understandings of a certain word, so in many cases, the content matched to the words specified by me may be different from others.}''.
}
Second, \shiwei{during the \textit{search} step,} seven participants also found it difficult to \textit{find posts that were similar to the explored one} (\textbf{C2}), \eg in terms of topics and sought social support, as indicated by seven participants. 
For example, S08 (M, 23) stated how he struggled to find another post with a similar topic to the current post. 
``\textit{I just explored a post about peer pressure in research and found it helpful. I returned to the search page and would like to check another related post that could offer more references for me to handle peer pressure. However, the titles of many returned posts queried by the keywords `peer pressure' and `research' could not tell whether the posts are related to the peer pressure in research}'' (S08). 
Third, \shiwei{seven participants found that searching with keywords could not help them \textit{filter comments that provided the needed type of social support} (\textbf{C3}). They had to read the content in detail to identify the specific type of social support, which hindered the convenience of the support-seeking process.}
The comments in OMHCs are of various quality regarding their provided informational and emotional support \cite{sharma2018mental}. 
Seven participants indicated that they frequently encounter comments that \proofreading{were} of low quality or that \proofreading{failed} to provide the type of social support they require. 
For instance, S10 (M, 25) stated, 
``\textit{My job search hasn't been going so well lately, so I'm trying to see if anyone else has any job search advice, but I've noticed that I keep seeing comforting words, but I'd rather see some useful advice.}''.

\textbf{Making sense of the helpful content in the Organize\&Digest steps takes a lot of mental effort.} 
When reading posts in detail, our participants expected to make sense of helpful user-generated content that comforted them or provided actionable suggestions.
\shiwei{They reported two challenges during the 
\proofreading{\textit{organize} and \textit{digest} step}
for information sensemaking.}
First, five participants indicated it was hard to \textit{organize useful content to 
\revision{cope with mental health concerns}
}
(\textbf{C4}) during the \textit{organize} step.
For example, S05 (M, 22) said, 
``\textit{I read several small suggestions across different posts and comments, but after the 30-minute exploring session, I could not form a clear plan for tackling my problem, as the suggestions are granular. I wish I could organize these suggestions when reading them effectively not in mind.}''.
\shiwei{
Second, six participants indicated they had difficulty in \textit{making sense of domain-specific content} (\textbf{C5}), which prevented them from digesting the useful content.
}
Six participants recalled that they 
\proofreading{read}
content that could be helpful but they needed more information to understand the content and make use of it. 
Some proper nouns may be \proofreading{confusing} for someone who lacks domain-specific knowledge.
S06 (M, 22) \proofreading{shared} an example, 
``\textit{I wanted to get some suggestions from peers in the community to handle insomnia. However, I was unfamiliar with many terms mentioned in the insomnia-related posts. I had to switch to another web page to google them, which is inconvenient.}''. 
S10 (M, 25) added, 
``\textit{While I appreciate the value of the current proposal, I'm not entirely convinced it's suitable for my specific problem. 
I am wondering if it could provide some additional guidance on how this advice could be applied to my situation. I believe this would encourage deeper thinking and better digestion of the advice I have received.}''. 

\subsection{Design Requirements}
\shiwei{
We asked participants for suggestions on any added feature to \zhenhui{the r/Anxiety interface} to help them browse the community for support-seeking during the four steps (\ie search, read, organize and digest).
Our participants actively offered suggestions that may help them find desired posts and comments and make sense of the helpful content to 
\revision{cope with mental health concerns}.
}
Based on their suggestions and related work, we derived four design requirements (DRs) of \name{}. 

\zhenhui{
\textbf{DR1. Support visual exploration of posts to help viewers find posts related to their mental health concerns.}
}
In our formative study, participants found it difficult to articulate their mental health issues in keywords \textbf{(C1)}, especially when they did not have a clear mind at the beginning.
They expected to be able to find the desired post more intuitively, for example, based on similarity \textbf{(C2)}, as seven participants denoted.
Previous studies show topic visualization \cite{peltonen2017topic, tu2022phrasemap} can help users understand the main themes of textual data without reading all of them \cite{diakopoulos2014identifying}, which can accelerate the information-seeking process.
\zhenhui{
Thus, \name{} should support visual exploration of posts with features like topic clusters that help viewers interactively find posts related to their mental health concerns. 
}

\zhenhui{
\textbf{DR2. Support navigation of comments 
\revision{based on its conveyed social support}
to help viewers access needed social support.}
}
In our formative study, we found that it is difficult for participants to find their desired posts and comments to 
\revision{cope with their mental health concerns}.
Some participants preferred practical suggestions to overcome their problems, while others may value emotional support that makes them confident in themselves.
As mentioned by seven participants, they had difficulty in filtering comments that provided the needed type of social support \textbf{(C3)}.
Thus, \name{} should support the navigation of comments \revision{based on its conveyed social support} to help viewers access needed social support.

\zhenhui{
\textbf{DR3. Support interactive note taking of useful content to organize social support information from the community.} 
\shiwei{
In our formative study, participants were frequently required to engage in iterative searching, reading, collection, and organization of a substantial amount of information, including unstructured comments, which makes the support-seeking process a messy and overwhelming experience \cite{liu2022crystalline}.
}
For viewers with mental health concerns, it is necessary to provide appropriate, actionable solutions to assist in 
\revision{coping with their mental health concerns}, as indicated by all participants.
``\textit{I appreciate the comforting words, but I am looking for more tangible solutions to the problem that's causing me stress. I would prefer a tool that helps me organize a good solution based on the suggestions I have collected}'' (S04, M, 23).
}
However, mining valuable information from OMHCs is very challenging due to the unstructured and unorganized nature of the comments \cite{peng2023designquizzer, 10.1145/3555555}.
Yet, presenting the data source \zhenhui{in an organized way} can mitigate the challenge of digesting such information \cite{liu2023coargue}.
Four participants indicated that they needed to memorize and consider the feasibility of these suggestions while reading the posts. 
\shiwei{
Such a process can result in an increased cognitive load, particularly when the individual is experiencing anxiety.
}
During the experiment, \zhenhui{in a word document}, S09 (F, 23) \proofreading{employed} a variety of colors to differentiate between the suggestions that she viewed during the experiment. 
For example, green was used to indicate suggestions she agreed with, yellow to indicate suggestions she could try, and red to indicate suggestions she strongly disagreed with.
``\textit{I think organizing the information with different colors while reading makes it clear and easy to identify key points.}'' \textbf{(C4)}.
\zhenhui{Therefore, \name{} should provide an easy way for viewers to take notes and organize the useful social support information in the community.}

\zhenhui{
\textbf{DR4. Support sense making of domain-specific content encountered in the community.}
Six participants indicated that providing pertinent information about their current reading information to help them make sense of user-generated content would be beneficial in enabling them to apply the suggestions to their own mental health issues \textbf{(C5)}.
In our formative study, we found that participants had to switch to another web page to search some domain-specific content (\eg ACT, ASMR) which prevented them from fully engaging with the content.
``\textit{It annoys me that I can't find some straightforward advice in simple terms, rather than having to go to different sites to understand it}''(S06, M, 22).
``\textit{It would have been helpful to have some guidance during the reading. I see some suggestions that interest me, but I'm not sure how to specifically implement them to help me with my mental health issues}'' (S01, F, 22).
Therefore, \name{} should support viewers to make sense of domain-specific content whenever they meet in the community. 
}
\section{ComViewer}
\begin{figure}[h]
  \includegraphics [width=\textwidth]{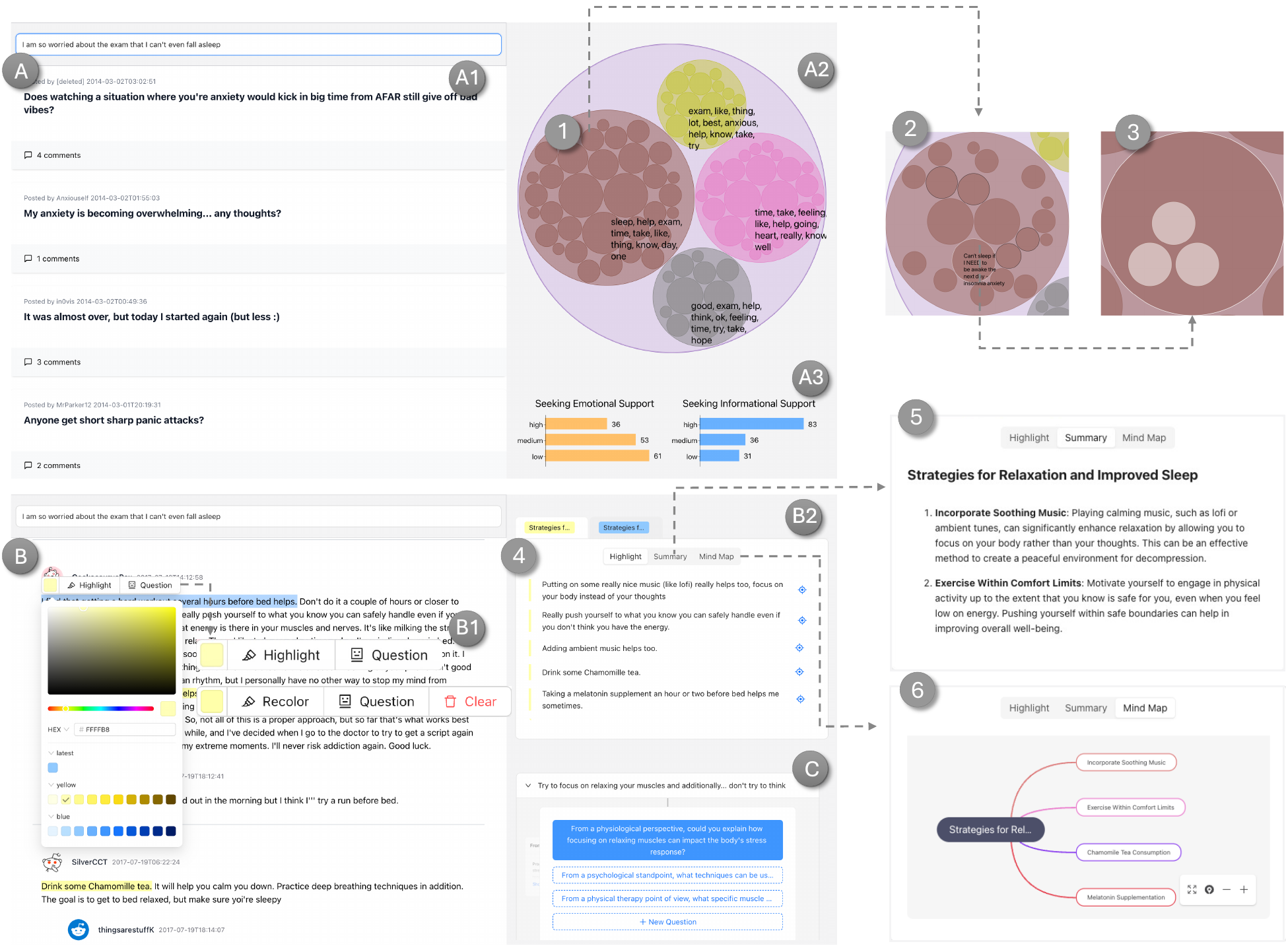}
  \caption{
  ComViewer: (A) The Zoomable Posts panel consists of (A1) the Post-list View, which shows the searching posts lists results in the OMHC; (A2) the Circle Packing View which allows viewers to interact with posts and comments with (1-3): three levels of posts visualization, and (A3) the Support Filtering View, which help viewers filter posts/comments based on their desired level of social support;  (B) The Note-taking panel consists of (B1) the Post-detail view with the interactive component which allow viewers to highlight or select reading content to inform organization and questioning, (B2) the Folder view which allows viewers to organize useful information while reading consists of (4-6) three folders according to colors; (C) the Questioning panel which allows viewers to ask questions while reading and displays the question flow.
  }
  \label{fig:all}
\end{figure}
Based on the design requirements derived from the formative study, we designed and developed \name{} (\autoref{fig:all}), an interactive visual tool that facilitates viewers to seek social support from posts and comments in online mental health communities (OMHCs). 
\autoref{fig:pro_pic} gives an overview of how \name{} with three panels, \ie the \vis{} panel, the \note{} panel, and the \question{} panel, supports viewers to search, read, organize, and digest information in OMHCs. 
In this section, we first describe a user scenario of \name{} and then detail the interface and implementation of each panel in \name{}.

\subsection{User Scenario}
This scenario describes Alice, an undergraduate student who is experiencing anxiety due to an upcoming exam.
She turns to an OMHC equipped with \name{} to browse posts and comments for suggestions to 
\revision{cope with her mental health concerns}.

\begin{figure}[]
  \centering
  \includegraphics[width=\linewidth]{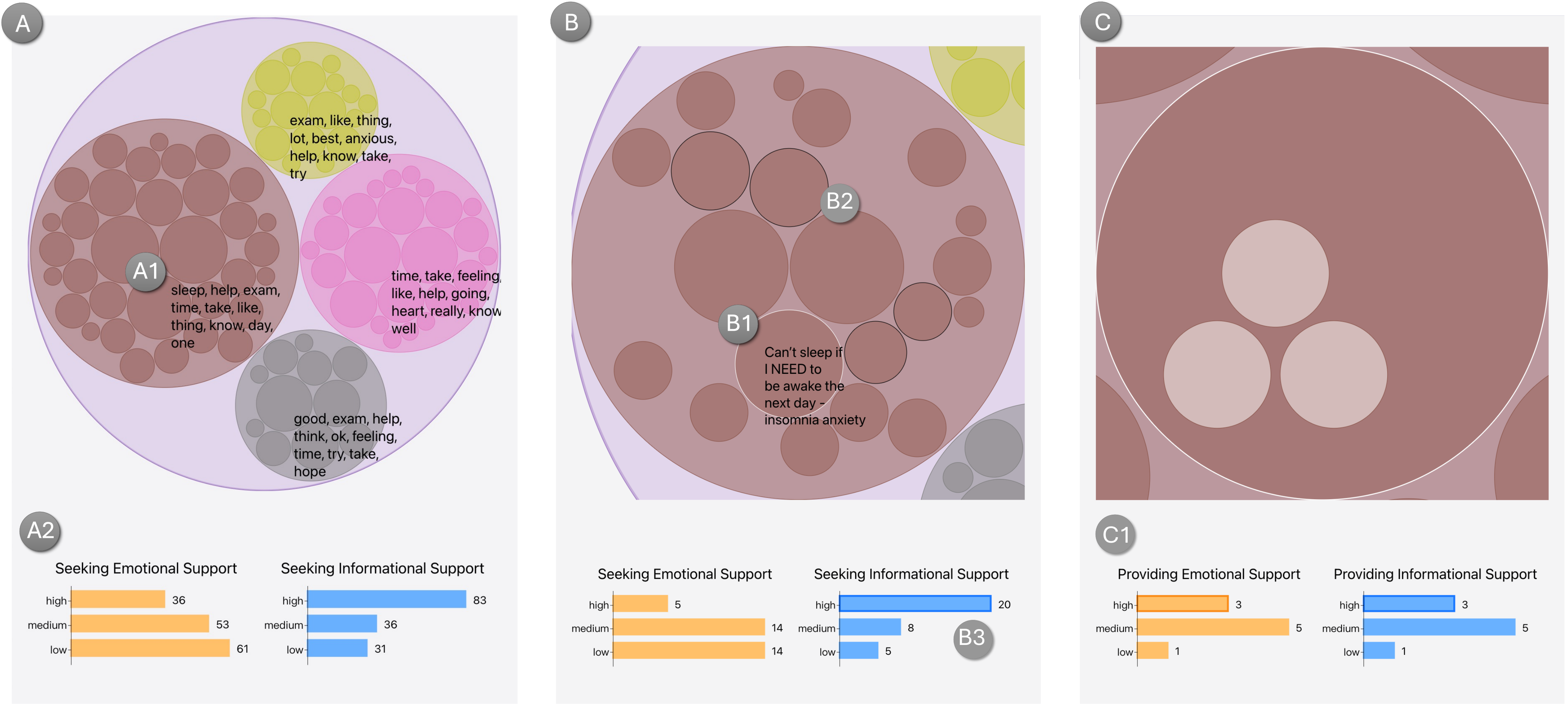}
  \caption{The Zoomable Posts Panel has three levels: (A) the topic level, (B) the post level, and (C) the comment level. (A1) Topical keywords for the posts cluster that are highly relevant to the search query. (A2) 
  \revision{Bar charts in which the length of and the number next to each bar indicate how many posts in the current circle pack seek a certain level of informational or emotional support in the topic or post level, or how many comments provide a specific level of informational or emotional support in the comment level.}
  (B1) A hovered post circle with a white border and its title displayed alongside. (B2) Posts with content similar to that of the currently hovered post are highlighted in black. (B3) Selecting the bar corresponding to posts that provide high informational support to help filter posts in the circle packing view. (C1) Selecting the bar corresponding to comments that provide high emotional and informational support to help filter \proofreading{comments} in the circle packing view.}
  \label{fig:C1_1}
\end{figure}

Alice first enters: ``I am so worried about the exam that I can't even fall asleep'' in the search area in the \vis{} panel (\autoref{fig:all} A).
She gets a list of returned posts and notices a set of packing circles that are highly relevant to the search query.
The following support filtering view (\autoref{fig:all} A3) illustrates the distribution of social support for the posts included in the circle packing view (\autoref{fig:all} A2) displayed above.
She selects a cluster in the topic level (\autoref{fig:C1_1} A) according to the keywords above the circle and then she zooms in to the post level (\autoref{fig:C1_1} B).
As she only wants to get some practical advice \zhenhui{rather than} comforting words from the community, she clicks the ``Seeking Informational Support - high'' bar to filter posts that attempt to seek high informational support (\autoref{fig:C1_1} B3). 
She hovers on several circles to read the titles of posts and clicks the largest one which means it has the most comments and then she zooms in to the comment level (\autoref{fig:C1_1} C).
At the same time, she navigates to the post-detail view (\autoref{fig:all} B1).
She reads the post content to determine if the experience described is similar to her own.
After zooming in and zooming out through the circle packing view (\autoref{fig:all} A2), Alice could find her desired post and read the comment to find some useful suggestions to help her 
\revision{cope with her mental health concerns}.
This time, she wants comments that would provide both high emotional and informational support, so she clicks both the ``Providing Emotional Support - high'' and ``Providing Informational Support - high'' bars to filter related comments (\autoref{fig:C1_1} C1).
She then clicks on one of the remaining comment circles.
Concurrently, the post-detail view directs her to the location of that selected comment.
While reading the comments, she identifies sentences that are particularly useful and highlights them in yellow. 
She also employs a different color to mark other points to distinguish the information.
Next, Alice clicks on the post circle to zoom out to the post level and then clicks on other similar posts and reads the comments. 
While reading, Alice encounters some suggestions she does not understand, so she selects the suggestion and clicks the pop-up ``Question'' button, which creates a folding board with recommended questions on the selected content in the \question{} panel (\autoref{fig:all} C). 
She clicks on one recommended question on the board and reads the answer. 
She also asks another question in response to the current answer (\autoref{fig:question} E). 
After the posts and comments exploration, Alice opens the yellow folder in the folder view (\autoref{fig:folder} A), which lists the content she highlights in yellow. 
Subsequently, the button (\autoref{fig:folder} A4) following the sentence is clicked to navigate to the designated highlighted position. 
This facilitates a comprehensive review of the suggestions, contextualized within the original content. 
Furthermore, she is able to adapt the suggestions by directly modifying them within the folder.
She then clicks the ``Summary'' button to view a machine-generated summary (\autoref{fig:folder} B).  

After visual searching, interactive organizing, and making sense of the content in the OMHC, Alice has a clearer understanding of how to 
\revision{cope with}
her mentally challenging problem. 
\subsection{Interface}
\name{} consists of three main panels: the \vis{} panel (\autoref{fig:all} A), the \note{} panel (\autoref{fig:all} B), and the \question{} panel (\autoref{fig:all} C).
\subsubsection{The Zoomable Posts panel}
The \vis{} panel provides an overview of search results and supports an interactive exploration of posts through zooming in and out among the topic, post and comment level, which enables viewers to identify and articulate specific mental health issues (\textbf{DR1}).
It also supports viewers to filter needed social support to help 
\revision{cope with their mental health concerns}
(\textbf{DR2}).

\textbf{Post-list view} (\autoref{fig:all} A1).
The post-list view displays the returned posts based on the searched keywords input in the search box like the traditional post-list layout.

\zhenhui{
\textbf{Circle packing view} (\autoref{fig:all} A2).
The circle packing view organizes the returned posts related to the searched keywords in a visual, interactive, and hierarchical way. 
\shiwei{It supports the \textit{search} step and the \textit{read} step in \autoref{fig:pro_pic} during which viewers face three challenges \ie C1, C2, and C3.
}
The first level of circles noted as topic-level circle packing (\autoref{fig:C1_1} A), enables viewers to observe the clusters of returned posts based on their topics. 
The circle's size is positively related to the number of posts included in the cluster. 
The keywords (\autoref{fig:C1_1} A1) on top of each circle reflect the topic of the clustered posts within the circle. 
Viewers can click a topic-level circle to zoom in to the second level of circles, noted as post-level circle packing (\autoref{fig:C1_1} B), in which each circle encodes a post that falls in the topic cluster. 
This time, the size of the circle is positively related to the number of comments that the post contains.
The circles in this level are colored consistently to enhance the visibility of packing.
Upon hovering over a given circle, other circles whose represented posts are similar to the hovered ones are marked with a black border within the current post cluster. 
This could facilitate users to locate similar posts (\textbf{DR1}). 
Viewers can further click on the post-level circle to zoom in \proofreading{to}
the third level of clusters, noted as comment-level circle packing (\autoref{fig:C1_1} C), and at the same time, the post-detail view will display the post associated with the circle and its comments. 
In the comment-level packing, each circle represents a comment of the same post, and a click on a circle will redirect viewers to the corresponding comment's location in the post-detail view (\autoref{fig:all} B1).
As viewers navigate between the three levels of the circle packing view, the lists of posts in the post-list view are updated in accordance with the posts contained in the different levels.
Viewers can either explore in the post-list view or by hovering over the circle with its title displayed adjacent to the selected circle (\autoref{fig:C1_1} B1).
}

\textbf{Support filtering view} (\autoref{fig:all} A3). 
The support filtering view under the circle packing view aims to help viewers filter the desired type and level of support \textbf{(DR2)}.
\shiwei{It supports the \textit{search} step by overcoming one challenge, \ie C3.}
\revision{The support filtering view is composed of bar charts, where the length of and the number next to each bar represent how many posts in the current circle pack seek a specific level of informational or emotional support, or how many comments provide a specific level of such support. This visual representation offers an overview of the social support at the three levels (\ie topic, post and comment level).
This distribution allows users to understand the social support sought across posts under specific keywords at the topic level, within groups of posts at the cluster level, and the social support provided by comments under individual posts at the comment level.
}
\zhenhui{
If the circle packing view is at the topic level or \proofreading{post} level, the two histograms are ``Seeking Emotional Support'' and ``Seeking Informational Support'', which encode the numbers of posts that seek different levels (high, medium, and low) of emotional and informational support. 
If it is at the comment level, the histograms turn to ``Providing Emotional Support'' and ``Providing Informational Support'', encode the numbers of comments under the post that provide different levels (high, medium, and low) of emotional and informational support. 
When a bar is clicked, its edge will be highlighted, and only the posts/comments that meet this condition are shown in the circle packing view (\autoref{fig:C1_1} B, C), which could help viewers directly find needed type of social support sought in the post or provided in the comments for 
\revision{coping with their mental health concerns}
(\textbf{DR2}).
}
\begin{figure}[]
  \centering
  \includegraphics[width=\linewidth]{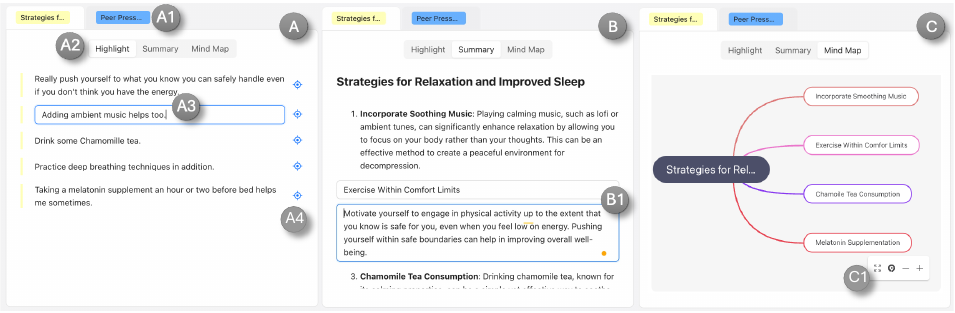}
  \caption{Folder view: (A) Collection page (B) Summary page (C) Mind map page.
  (A1) Folder labels with the same color as the highlighted color for color-coded content organization. (A2) Three on-page conversion buttons. (A3) Modifications to the collection. (A4) Button to jump back to the highlighted address in the text. (B1) Modifications to the summary. (C1) Operations component for mind map.}
  \label{fig:folder}
\end{figure}
\subsubsection{Note-taking panel}
The \note{} panel aims to assist viewers in organizing social support structurally to reduce the mental effort required while consuming the suggestions in the OMHC (\textbf{DR3}).
\shiwei{It supports the \textit{organize} and \textit{digest} step by overcoming the encountered challenge \ie C4.}

\zhenhui{
\textbf{Post-detail view} (\autoref{fig:all} B1).
The post-detail view displays the content of a post and its comments.
When reading the content, viewers can highlight or select any text to collect, summarize, and question it.
They can highlight the content with different colors, remove it by clicking the ``Clear'' button, and recolor it with the ``Recolor'' button.
}

\textbf{Folder view} (\autoref{fig:all} B2).
The folder view 
consists of three pages: \textit{Collection} page (\autoref{fig:all} (4)), \textit{Summary} page (\autoref{fig:all} (5)), and \textit{Mind Map} page (\autoref{fig:all} (6)).
The highlighted content can be automatically integrated into folders with the same color label (\autoref{fig:folder} A1) in the \textit{Collection} page \textbf{(DR3)}.
At the same time, a machine-generated summary based on the collected information will be displayed in the \textit{Summary} page every time the highlighted content is updated.
This allows viewers to conveniently review the suggestions with a more refined categorization, which could reduce their cognitive load of organizing the suggestions.
Additionally, a mind map reflecting subtitles of the summary, indicating key points of the suggestions is generated on the \textit{Mind Map} page.
Viewers can utilize this map according to their needs to facilitate the flexible organization of useful suggestions.
To facilitate customized data exploration, we allow viewers to edit the collected information in the \textit{Collection} page (\autoref{fig:folder} A3), the generated summary in the \textit{Summary} page (\autoref{fig:folder} B1), and the mind map in the \textit{Mind Map page} (\autoref{fig:folder} C) according to their preferences.
Viewers can navigate to the corresponding highlighted area by utilizing the navigation button (\autoref{fig:folder} A4) to recall the suggestions by reading the corresponding context.
\begin{figure}[h]
  \centering
  \includegraphics[width=0.8\linewidth]{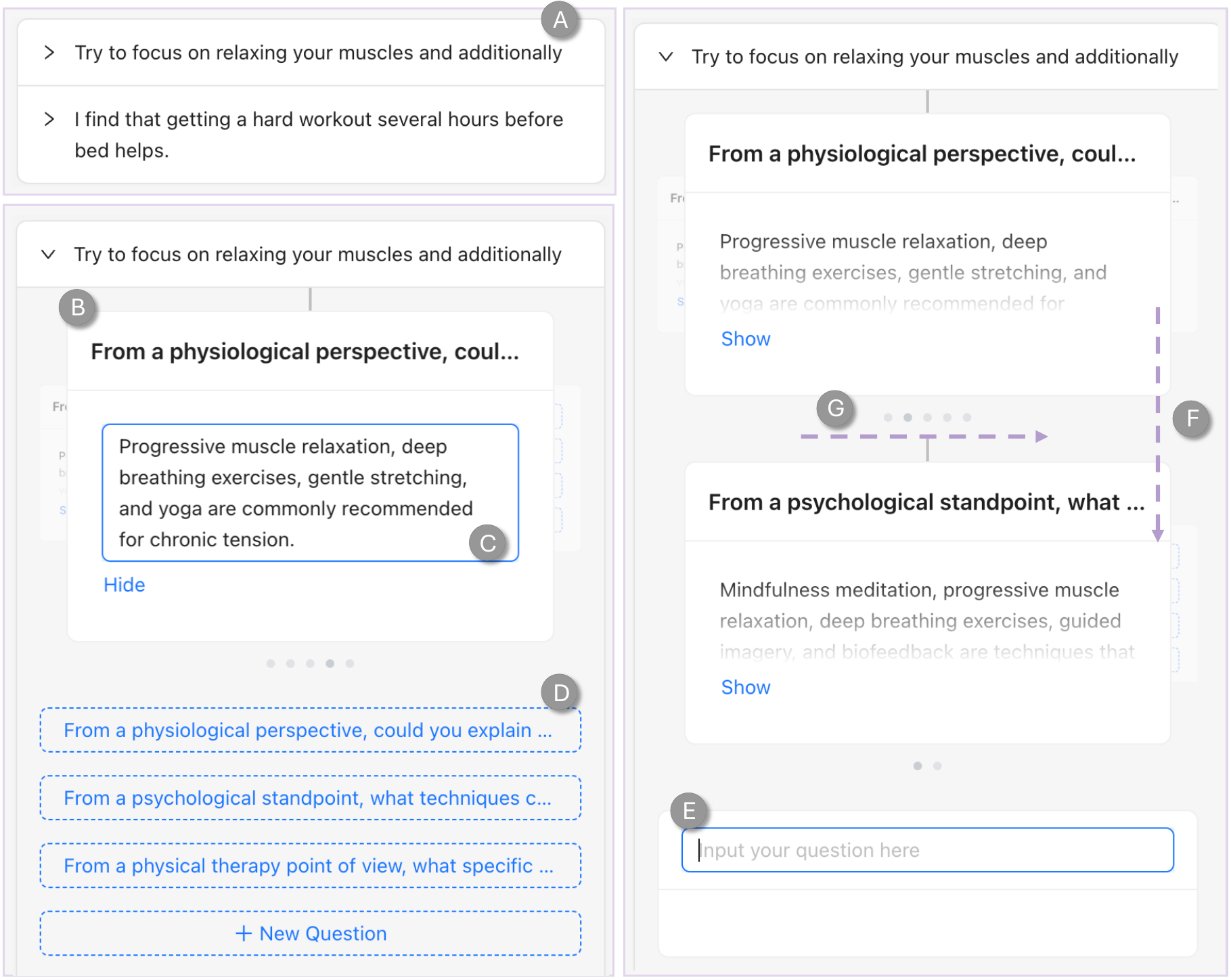}
  \caption{Questioning Panel: (A) Folder boards to show the list of questions. (B) Whiteboard with a question and generated answer. (C) Modification to the answer. (D) Recommended questions.  (E) Input box for adding new questions. (F) Linear layout of whiteboards to show continuity of questions. (G) Horizontal layout of questions to demonstrate multi-dimensionality of questions.}
  \label{fig:question}
\end{figure}
\subsubsection{Questioning panel} 
The \question{} panel assists viewers in making sense of domain-specific content by providing recommended questions to the selected content and answers to the questions to 
\revision{help cope with individual mental health concerns}
(\textbf{DR4}).
\shiwei{It overcomes the encountered challenge in the \textit{digest} step, \ie C5.}
It supports viewers to ask questions about \zhenhui{any textual content} they are currently reading to get additional information to fulfill their information needs. 
\zhenhui{
To use the \question{} panel, viewers should first select content in the post-detail view and click the pop-up ``Question'' button (\autoref{fig:all} B1), which will generate a folding board (\autoref{fig:question} B) with the selected content displayed at the top (\autoref{fig:question} A), followed by three recommended questions (\autoref{fig:question} D) generated by a large language model. 
Viewers can then click any of the recommended questions or ask their own questions, which will generate a whiteboard (\autoref{fig:question} B) with the question on the top and the answer below.
\shiwei{
The system will suggest questions based on previous responses, thereby assisting viewers in progressing through the process.
}
Besides, viewers can edit the answer (\autoref{fig:question} C) according to their preference which may result in a different impact on the recommended questions generated based on the answer.
Viewers may observe the ongoing question process for a selected content item (\autoref{fig:question} F), as well as multiple perspectives on that item (\autoref{fig:question} G). 
Furthermore, the details about the question's progress can be hidden using the buttons (\autoref{fig:question} A) on the front of the folder board with only the selected item left. 
Upon clicking on the grey dots (\autoref{fig:question} G) situated beneath the pertinent whiteboard a new conversation in parallel is generated for viewers to pose another question in accordance with the previous query.
These features could facilitate viewers to continue with the previous questioning processes to have a deep exploration of the selected content. 
}
\subsection{Implementation Detail}
\name{} is implemented as a web-based application using Python and a Flask backend with React frontend. 
It is deployed on a web server so that users can easily visit \name{} in their browsers. 
We do not implement it as an extension or plug-in to existing OMHCs as we do not have permission to access the built-in APIs and modify (\eg highlight, sort) the content of these communities.
We leverage Elasticsearch \cite{elasticsearch}, which helps \name{} scale to a large number of data repositories for indexing and search. 

\subsubsection{Data} \label{sec:data}
\textbf{Data Collection}.
To demonstrate \name{}, we select the data from Reddit r/Anxiety \zhenhui{(\autoref{sec:research_site})}.  
\zhenhui{We collect publicly available posts and comments created between Jan 2009 to Mar 2019 in this OMHC via Pushshift API \cite{reddit} in 2019.}
We take two steps to pre-process the collected data. 
First, we remove the posts and comments whose postid are ``[deleted]'' or ``[removed]'' since we cannot track their relevant comments or which posts they belong to.
Then, we remove the posts and comments whose content is ``[deleted]'' or ``[removed]''.
After pre-processing, our dataset consists of $182,880$ posts and $910,071$ comments.

\textbf{Labeling seeking/providing emotional and informational support}. 
We use the models \zhenhui{open-sourced by} \cite{10.1145/3411764.3445446} to label the amount of sought emotional support (ES) and informational support (IS) in the posts regarding three levels: high, medium, and low. 
The models for labeling sought ES/IS are MLP/XGBoost models and were trained by \cite{10.1145/3411764.3445446} (accuracy = 65/70\%) on an annotated dataset of Reddit r/depression, which is the same type of OMHCs as r/Anxiety \cite{sharma2018mental}.
Similarly, we use the models in \cite{peng2020exploring} to label the amount of provided social support in the comments regarding three levels: high, medium, and low. 
The models for labeling provided ES/IS are XGBoost/Random Forest models and were also trained (accuracy = 68/64\%) on an annotated dataset of r/depression. 
\zhenhui{
The coding criteria of sought social support can refer to the link \footnote{\url{https://zhenhuipeng.com/projects/satisfaction_mental_health/Sought_support_rating_scheme.pdf}} shared by \cite{10.1145/3411764.3445446}. 
}

\textbf{Data encoding}. 
We use the ``all-MiniLM-L6-v2'' model \footnote{\url{https://huggingface.co/sentence-transformers/all-MiniLM-L6-v2}} to encode the content of each post into a vector, calculate the similarity using cosine similarity and save the post pairs whose similarity are above a threshold (0.6 in our case) into Elasticsearch. 

\subsubsection{Zoomable Posts panel} \label{sec:LDA}
Upon inputting keywords into the search area, \name{} will automatically use Elasticsearch to search for relevant posts, which employs an inverted index and the TF-IDF algorithm for matching and returns the results in a sorted order based on the relevance of the search query, similar to the traditional OMHC searching algorithm.
Then, the circle packing view employs Latent Dirichlet Allocation (LDA) \cite{blei2003latent} and sets the number of topics to four in order to cluster the 150 posts with the highest relevant scores to the search query. 
Five sets of keywords are randomly selected and used to construct LDA models with the number of themes ranging from one to ten for the search results.
The mean number of themes that corresponded to the maximum coherence score is four.
This is because we find that when the keyword descriptions are highly detailed, the search results become highly homogenized and the number of clusters should be limited. 
Instead, the number of clusters should be refined.
The primary principle of our design is to leverage and enhance familiar visual metaphors.
We adhere to the guiding principle of ``overview first, zoom and filter, then details-on-demand'' \cite{shneiderman2003eyes} for users to actively explore posts, thereby enhancing their support-seeking experience. 
\zhenhui{
First, for the \textit{Overview} principle, the \vis{} panel shows an at-a-glance overview of post distribution with cluster keywords.
Second, for the \textit{Zoom and Filter} principle, the Circle Packing view allows viewers to zoom in and zoom out among the three levels (\ie topic, post, and comment level). 
The Support Filtering view allows viewers to select the needed type and level of social support.
Lastly, for the \textit{Details-on-Demand} principle, hovering on the circle can show the post title and similar posts.
}

We use D3.js \footnote{\url{https://d3js.org/}} to implement the Circle Packing and Support Filtering view in the \vis{} panel.
\mingxiang{With the data of three levels, we first reorganize it according to the parent-child hierarchy, and then plot nested circles of different sizes which represent the corresponding levels. 
To support zooming in and zooming out, we bind mouse events to each circle, which displays the corresponding content (\ie post title, topic keywords) when the mouse hovers over the circle, zooms in to a more detailed level when clicking inside the circle, and zooms out to the more abstract when clicking outside. 
Support Filtering view consists of two bar charts, which represent the distribution of emotional support and informational support. 
With the component communication mechanism provided by React, the click event bound to the bars can filter the data of circle packing according to the corresponding types and levels of social support.}

\subsubsection{Note-taking and Questioning panel} 
We leverage LangChain \footnote{\url{https://github.com/langchain-ai/langchain}}, a framework that enables LLMs to analyze lengthy inputs.
We use the OpenAI's `gpt-4' model as our LLM.
As for the recommended questions, we design the prompt to generate questions that start with ``what'', ``why'', and ``how to do'' inspired by \cite{10.1145/3586182.3615785} from the viewers with mental health issues perspectives.
\shiwei{
As for the generated answers, we request the LLM to consider answering the questions from the perspective of a mental health professional.
}
The prompts used are shown in \autoref{sec:appendix}.

\mingxiang{
To support note-taking and questioning, we structure and store persistently the highlighted and questioned content in the form of JSON, which are shared among all components in two panels as global states and managed by Redux \footnote{\url{https://redux.js.org/}}. 
By binding click events and using React Router \footnote{\url{https://reactrouter.com/en/main}} for page navigation, we link the notes taken by users to the corresponding posts displayed in the post-detail view, to locate the highlighted content. 
For the implementation of the highlight feature, we refer to the code \footnote{\url{https://github.com/alienzhou/web-highlighter}}.
For the implementation of the mind map, we refer to the code \footnote{\url{https://github.com/SSShooter/mind-elixir-core}}.
The two panels are developed based on the UI component libraries: Ant design \footnote{\url{https://github.com/ant-design/ant-design}} and Arco design \footnote{\url{https://github.com/arco-design/arco-design}}.} 
\shiwei{ Our code will be made available for open-source and encourage future researchers to further customize them \footnote{\url{https://github.com/ShionMing/ComViewer}}.}
\section{User Study}

\zhenhui{
To evaluate the effectiveness and user experience of \name{} for facilitating viewers to seek social support in online mental health communities (OMHCs), we conduct a within-subjects study with 20 participants. 
Our research questions in the user study are: 
\begin{itemize}
\item \textbf{RQ1}. How would \name{} affect the \textbf{outcome of their sought social support} after exploring the community?
\item \textbf{RQ2}. How would \name{} affect viewers' engagement and cognitive load in the \textbf{process of seeking social support }in the community?
\item \textbf{RQ3}. How would viewers i\textbf{nteract and perceive with \name{}}?
\end{itemize}

\subsection{Baseline Interface}
The baseline interface has the same Post-list View (\autoref{fig:all} A1) and a similar Post-detail view (\autoref{fig:all} B1) but without the highlighting and questioning features as \name{}. These two views simulate the traditional practice of searching and reading posts in OMHCs (\eg 7cups \footnote{\url{https://www.7cups.com}}, Reddit r/Anxiety (\autoref{sec:research_site})). 
To control the amount of information attained and the consistent search results for the same search query, both \name{} and the baseline interface use the same database collected from Reddit r/Anxiety (\autoref{sec:data}) and the search algorithm with Elasticsearch (\autoref{sec:LDA}).
Moreover, to minimize the potential interfaces of other factors, such as UI design style \cite{10.1145/3411764.3445325}, the baseline interface is implemented to resemble \name{} in UI components and styles.
The only difference with \name{} is that the baseline does not have our proposed features in the \vis{} panel, the \note{} panel, and the \question{} panel.
\zhenhui{We do not include crowd-contributed tags (\eg flairs for ``uplifting'', ``advice needed'' or ``needs a hug'') in both interfaces because, in our formative study, we find that participants rarely use these tags.
They think the tags are perceived as overly broad and lacking in specificity, particularly in relation to the search for content using flairs ``health'' or ``advice needed''. 
We also do not include features like commenting and upvotes in both interfaces as we target the viewers who do not directly participate in the conversations in OMHCs. 
We will discuss the trade-off of this design choice in the Limitations and Future Work
\proofreading{(\autoref{sec: limitation})}.
}
}

\subsection{Participants and Tasks}
We recruited 20 participants (13 females, 7 males; age: 20-30, $Mean = 22.7, SD= 1.72$) through online advertising and word-of-mouth at a local university. 
The inclusion criteria were that participants had recently encountered mentally challenging issues about relationships and exams, had experience in viewing posts in OMHCs (\eg support groups in TikTok or Weibo), and were interested in exploring OMHCs to seek social support to cope with these issues. 
Three participants reported that they used OMHCs daily, ten people used OMHCs 4-6 days a week, two people used OMHCs at least once a week, and five people seldom used OMHCs. 
All participants were fluent in English and had passed the National College English Test.
Each participant had two \zhenhui{support-seeking} tasks \peng{about seeking support to address stressful issues about relationships and exams, as stated in the inclusion criteria. We controlled the target stressful issues to allow statistical comparison of the results in \name{} and baseline conditions}. 
\zhenhui{Before the study, we asked participants to score their ability to cope with these two kinds of mental health issues using items
``\textit{I believe that I am well-equipped to deal with relationship issues/exam stress}'' on a standard 7-point Likert scale, with 1 for strongly disagree and 7 for strongly agree.
The mean self-assessment of the ability to cope with relationship/examination stress is $4.15 (SD = 1.62) / 3.75 (SD = 1.89)$, which indicated that they were not quite good at coping with these mental health issues.}
The task prompts are: 

Task A (relationships): One of your friends said something bad to you. It made you hurt. How would you find helpful suggestions and comforting words and make sense of them to 
\revision{cope with}
your mental health concerns?

Task B (exam): Exams are approaching. You feel so anxious that you can not sleep. How would you find helpful suggestions and comforting words and make sense of them to 
\revision{cope with}
your mental health concerns?

We counterbalanced the tasks and interfaces using Latin Square. 
There are four task-interface combinations, each with five randomly assigned participants: (a) task A (\name{}) - task B (Baseline), (b) task B (\name{}) - task A (Baseline), (c) task A (Baseline) - task B (\name{}) and (d) task B (Baseline) - task A (\name{}).

\subsection{Measures}
\textbf{RQ1. Support-seeking outcome}.
In each task, we asked participants to summarize useful suggestions or comforting words that they thought could help 
\revision{cope with the mental health concerns}
in a document while exploring the OMHC for each task.
We \proofreading{counted} the number of meaningful points in each of these documents regarding informational support (\eg advice, knowledge) and emotional support (\eg comfort, encouragement), which are two common types of social support exchanged in OMHCs \cite{10.1145/3411764.3445446, peng2020exploring}. 
Two authors independently coded the meaningful points of twenty randomly sampled summary documents and then met and discussed their codes. 
For example, ``\textit{If you really care about your friend and are determined to keep her in your future, go talk to her and get clarity and peace of mind.}'' can be a meaningful point about informational support. 
``\textit{Misunderstandings and conflicts between friends are common, and they may not be well thought out or intended to be hurtful.}'' is considered a meaningful point about emotional support. 
Others like ``\textit{Have your own solutions for things that need to be addressed so that you can really get relief!}'' are considered meaningless.
After reaching an agreement on the coding scheme, they applied it to the rest outcome documents in a shuffled order and resolved the disagreement via discussions.
\revision{The researchers primarily focused on categorizing and organizing the points, splitting or merging them to ensure more accurate statistics, and made their judgments based on participants' previous sharing to ensure that the coded meaningful points align with participants' perspectives.}
Besides, after each task, we measured their perceived satisfaction and confidence with their received social support from the OMHC using two items adapted from \cite{10.1145/3555555}, \ie ``I am satisfied with the received social support from the OMHC about \textit{relationships} (task A) / \textit{exam} (task B)'' and %
``I am confident that the received social support from the OMHC can help me alleviate mental health issues about \textit{relationships} (task A) / \textit{exam} (task B)''. 

\textbf{RQ2. Support-seeking process}.
We adapted four items from \cite{10.1145/3555555} to measure user engagement in the support-seeking process. 
They are concentration (``completely involved, focused, and concentrating''), a sense of ecstasy (``process unique compared to previous ones''), clarity (``obtain a precise understanding of the current situation''), and doability (``skills are adequate, neither anxious nor bored'').
As for the cognitive load in searching desired posts and comments and making sense of the information, we \proofreading{used} two items adapted from \cite{peng2023designquizzer}: ``I think the cognitive load required to \textit{find the desired posts and comments} / \textit{make sense of the information I read} using the tool is minimal''.  

\textbf{RQ3. Perceptions towards \name{}}.
In both \name{} and the baseline conditions, we asked participants to rate their perceptions towards the OMHC interface using three items adapted from the technology acceptance model \cite{venkatesh2008technology}.
They are usefulness (``The interface is useful for me to find useful posts and comments as well as reading and digesting the content''), easy to use (``I think the system easy to use''), and intention to use (``I think I would want to use the system again for similar mental health problems''). 
\zhenhui{
In the \name{} condition, we also asked participants to rate the usefulness of each unique feature (\autoref{tab:feature}) using the item ``I found [feature] useful in my support-seeking process in the OMHC''. 
}

All the self-reported items in RQ1-3 were rated on a standard 7-point Likert scale, with 1 for strongly disagree and 7 for strongly agree.

\subsection{Procedure}
\begin{figure}[h]
  \includegraphics [width=0.8\textwidth]{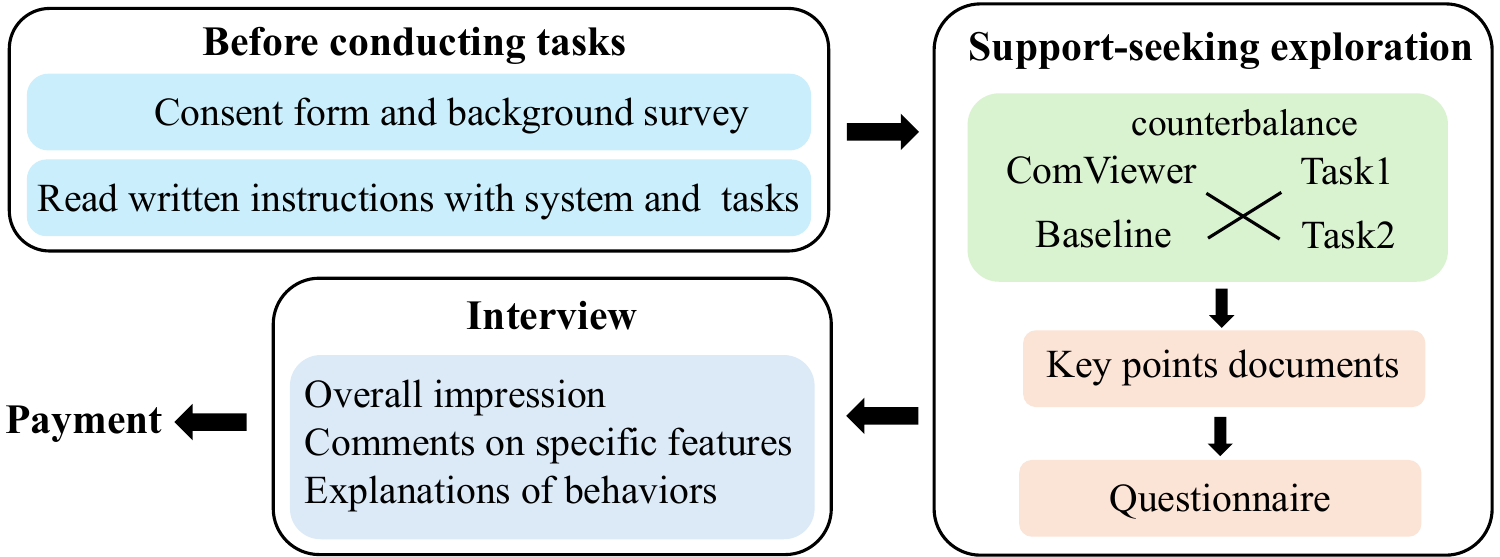}
  \caption{
  Procedure of the within-subjects user study. 
  In each information-seeking task, participants explore posts and comments with the assigned interface to seek social support.
  }
  \label{fig:userstudy}
\end{figure}
\zhenhui{
\autoref{fig:userstudy} illustrates the procedure of our user study.
We conducted the user study with each participant remotely via a video-conferencing application.
Before starting each task, we demonstrated the use of the basic functions of the assigned OMHC interface (\ie either \name{} or the baseline) and gave the participants 3-5 minutes to familiarize themselves with the interface. 
After that, we asked them to spend around 25 minutes exploring the posts and comments using the assigned interface to seek social support on the mentally challenging issue specified by the task. 
After each task, we asked participants to write down the key points learned from the community that could be useful for 
\revision{coping with their mental health concerns}.
\revision{Then, they were asked to share and discuss the key points they had noted.}
They also needed to fill in a questionnaire about their perceived emotional support from the community, engagement, and cognitive load in this support-seeking process, as well as their perceptions towards the used community interface. 
Upon completion of two tasks, we conducted a semi-structured interview with each participant, which focused on their overall impression of \name{}, comments on the specific features, and explanations of their behaviors when interacting with the systems. 
The user study lasted 60-90 minutes for each participant, and each participant received 60-90 RMB for compensation. 
}

\subsection{Data Analyses}
For the self-reported items in RQ1-3, \shiwei{which are ordinal variables}, we performed the Wilcoxon signed-rank test \cite{woolson2007wilcoxon} to assess the difference in the participants’ ratings between the \name{} and baseline conditions.
For the number of meaningful points, we first conducted a Shapiro-Wilk test \cite{shapiro1965analysis}.
Results showed that the informational support variable received a p-value greater than 0.05 (\name{}: $W(20) = 0.934, p = 0.186$; baseline: $W(20) = 0.951, p = 0.385$), which indicated the normal distribution of this item, 
while the emotional support variable indicated a significant departure from the normal distribution for \name{} ($W (20) = 0.773, p < 0.001$). 
We therefore conducted a paired sample t-test \cite{ross2017paired} to compare the differences of the number of key points about informational support and a Wilcoxon signed-rank test to compare the differences of the number of key points about emotional support between the \name{} and baseline conditions.
For the interview recordings, two of the authors transcribed them into text and conducted a thematic analysis \cite{braun2012thematic}. 
They first familiarized themselves by reviewing all the text scripts independently. 
After several rounds of coding with comparison and discussion, they finalized the codes of all the interview data. 
We counted the occurrences of codes and incorporated these qualitative findings in the following presentation of our results.

\section{Results}
\autoref{tab:userStudyResults} summarizes the statistical results of the measured items for RQ1-RQ3.
\begin{table}[htbp]
    \centering
     \caption{The experiment's RQ1-3 statistical results about ComViewer and the baseline OMHC interface. 
     Perceived information support was analyzed using paired sample t-test, while others using Wilcoxon signed
     rank test.
     All items except the two for RQ1 are measured using a standard 7-point Likert scale (1 - strongly disagree; 7-strongly agree). Note:$***:p <0.001;**:p<0.010;*:p<0.050;+:0.050<p<0.100;-:p>0.100$ ; within-subjects; N = 20.}
    \label{tab:userStudyResults}
    \begin{tabular}{clllllc}
    \toprule
    {Research} & 
    \multirow{2}{*}{{Item}} &
    {{ComViewer}} &
    {{Baseline}} &
    \multicolumn{3}{c}{Statistics}  \\
    \cline{5-7}
    {Question} & ~ & Mean (SD) & Mean (SD) & {$Z/T$} & {$p$} & $Sig.$ \\
    \hline
    \multirow{6}{*}{\shortstack{(RQ1) \\Support-seeking \\outcome}} &
     {Number of points about}   & \multirow{2}{*}{1.25 (1.55)} & \multirow{2}{*}{1.65 (1.23)}  & \multirow{2}{*}{1.198} & \multirow{2}{*}{0.231} & \multirow{2}{*}{-} \\
     ~ & {emotional support} & ~ & ~ & ~ & ~ & ~ \\
     ~ & {Number of points about} & \multirow{2}{*}{5.25 (2.25)} & \multirow{2}{*}{3.30 (1.87)} & \multirow{2}{*}{4.124} & \multirow{2}{*}{<0.001} & \multirow{2}{*}{***}\\
     ~ & {informational support} & ~ & ~ & ~ & ~ & ~ \\
     \cline{2-7}
     ~ & Satisfaction & 5.60 (1.10) & 4.40 (1.54)  &  -2.200 & 0.028 & * \\
     ~ & Confidence & 5.55 (0.94) &4.25 (1.45) & -2.978 &0.003 & ** 
    \\
    \hline
    \multirow{8}{*}{\shortstack{(RQ2) \\Support-seeking \\process}} &
    Concentration   & 5.70 (1.26) & 4.40 (1.44)  & -3.618 & 0.009 & ** \\
    ~ &  Sense of ecstasy & 6.35 (0.98) & 2.90 (1.58)  & -3.723 & <0.001 & *** \\
    ~ &  Clarity   & 5.60 (0.99) & 3.70 (1.71)  & -3.493 & <0.001 & *** \\
    ~ &  Doability & 5.65 (1.50) & 3.40 (1.60)  & -3.049 & 0.002 & ** \\
    \cline{2-7}
    ~ & {Cognitive load} & \multirow{2}{*}{5.80 (0.77)} & \multirow{2}{*}{3.20 (1.74)}  & \multirow{2}{*}{-3.614} & \multirow{2}{*}{<0.001} & \multirow{2}{*}{***} \\
    ~ & {for searching} & ~ & ~ & ~ & ~ & ~ \\
    ~ &  {Cognitive load} & \multirow{2}{*}{5.55 (1.05)} & \multirow{2}{*}{3.10 (1.52)}  & \multirow{2}{*}{-3.641} & \multirow{2}{*}{<0.001} & \multirow{2}{*}{***} \\
    ~ & {for sensemaking} & ~ & ~ & ~ & ~ & ~ \\
    \hline
    \multirow{3}{*}{\shortstack{(RQ3) Perceptions \\towards the \\interface}} &
    Usefulness  & 6.20 (0.52) & 3.05 (1.36) & -3.936 & <0.001 & *** \\
    ~ & Intention to use   & 5.95 (1.23) & 3.45 (1.57)  & -3.311 & < 0.001 & *** \\
    ~ & Easy to use & 5.45 (1.54) & 4.35 (1.79)  & -2.157 & 0.031 & * \\
    \bottomrule
    \end{tabular}
\end{table}

\begin{table}[htbp]
    \centering
    \caption{Participants' ratings on \zhenhui{the perceived usefulness of} \name{} features}
    \label{tab:feature}
    \begin{tabular}{llll}
    \toprule
     Component   &  Feature & Mean & SD \\
     \hline
     \multirow{5}{*}{\shortstack{Zoomable \\Posts panel}}
         & {Cluster returned posts based on their topics} & 5.80 & 1.06 \\
        ~ & {Provide an interactive posts exploration in a zoomable pattern} & 5.95 & 1.15 \\
        ~ & {Filter posts/comments based on} & \multirow{2}{*}{5.80} & \multirow{2}{*}{1.06} \\
        ~ & {the amount of sought/received support} & ~ & ~ \\
        ~ & {Recommend similar posts based on content} & 5.45 & 1.20 \\
    \hline
    \multirow{5}{*}{\shortstack{Note-taking \\panel}}
        & {Highlight any content using different color while reading} &6.30 &0.92 \\
        ~ & {Navigate to previous highlighted content} & 6.55 & 0.60 \\
        ~ & {Organize highlighted content of the same color} & 6.50 & 0.76 \\
        ~ & {Generate a summary of the highlighted content} & 6.15 &1.09 \\
        ~ & {Provide the summary in a mind map pattern} & 5.75 &1.33 \\
    \hline
    \multirow{2}{*}{\shortstack{Questioning \\panel}}
        & {Recommend questions from different perspectives} & 5.80 & 1.11 \\
        & {Interact with the questions and answers in a hierarchy layout} & 5.90 & 1.02 \\
    \bottomrule
         
    \end{tabular}
\end{table}

\subsection{Impact on Support-Seeking Outcome (RQ1)}
\label{sec:rq1_result}
There is no significant difference in the number of key points that participants record about emotional support between \name{} ($Mdn = 1.00$) and the baseline ($Mdn = 1.50$) conditions; $Z = 1.198, p = 0.231$.
However, participants with \name{} ($Mdn = 5.00$) record significantly more points learned from the online mental health community (OMHC) that provide meaningful informational support, compared to the conditions when they are with the baseline OMHC interface ($Mean = 3.30, SD = 1.87$); $T = 4.124, p < 0.001$. 
Overall, participants with \name{} ($Mean = 5.25, SD = 2.25$) are significantly more satisfied with the social support that they have received by exploring the OMHC than the received social support when they used the baseline interface ($Mdn = 5.00$); $Z = -2.200, p = 0.028$. 
Compared to the baseline condition ($Mdn = 4.00$), participants in the \name{} condition ($Mdn = 6.00$) are also significantly more confident that the received social support would help them address their mental health issues; $Z = -2.978, p = 0.003$. 

These results suggest that \name{} helps participants get significantly more advice or information from the OMHC, and they are more confident that informational support would help them cope with their mentally challenging issues. 
These results could be accounted for the simplified process of searching and making sense of useful content supported by \name{}, as agreed by all participants. 
For example, P8 (F, 23) stated the efficiency of using the \vis{} panel to find useful content,  
``\textit{I can zoom in and zoom out the circles to explore the posts based on their topics and the types of sought support and explore the comments that provide informational support. It really reduced my time spent on filtering useful content in the community}''.
P5 (F, 23) mentioned the convenience of \name{}'s highlighting and summarization features,  
``\textit{I can highlight the content that may be helpful and get an automatically generated summary on all the highlighted content. This helped me easily summarize the take-away advice learned from the community}''.
P9 (F, 22) talked about how the questioning feature helps her make sense of the comments that could be useful,  
``\textit{I was confused with a comment and wanted to learn more about it. One of the recommended questions about this comment is exactly what I am interested in. I made sense of this comment after checking the generated answers on this question}''.

\subsection{Impact on Support-Seeking Process (RQ2)}
\label{sec:rq2_result}
As shown in \autoref{tab:userStudyResults}, compared to the baseline interface \proofreading{($Mdn = 5.00, 2.50, 3.00, 3.00$)}, 
\name{} \proofreading{($Mdn = 6.00, 7.00, 6.00, 6.00$)} significantly improves participants' concentration ($Z = -3.618, p = 0.009$), sense of ecstasy ($Z = -3.723, p < 0.001$), clarity ($Z = -3.493, p<0.001$), and doability ($Z = -3.049, p = 0.002$) during the seeking-support process. 
\name{} ($Mdn = 6.00, 5.50$) also significantly reduces participants' perceived cognitive load in searching for useful information ($Z = -3.614, p < 0.001$) and sensemaking ($Z = -3.641, p < 0.001$) compared with the baseline system ($Mdn = 3.00, 3.50$).
These results indicate that participants are significantly more engaging and less mentally demanding in the process of exploring the OMHC with \name{} to seek social support.

The improved experience in the support-seeking process could be accounted for the unique features of \name{}. 
For example, nine participants highlighted that \name{}'s interactive \vis{} panel helped clarify their thoughts on their desired posts and comments. 
P11 (F, 22) stated, 
``\textit{
The graphs (\ie the circle packing view) helped me view the search results from a clear and comprehensive perspective, making it easier and more interesting to find helpful posts. 
In the other interface (\ie the baseline one), I had to look at each returned post, click it to check in detail, and return back to the searched results if it was not helpful. Such a process disappointed me}''.
P7 (M, 24) added, ``\textit{The visual clusters of the posts helped me quickly find relevant information to my situation}''.
Seventeen participants mentioned how the highlighting and summarization features eased their cognitive workload. 
For example, P9 (F, 22) said, 
``\textit{I can bookmark any content in \name{} and get a summary on it, while in the other interface, I have to paste the useful suggestions to the document which distracts my reading process, and spend effort in summarizing the takeaway based on these points}''.
P17 (F, 22) added, ``\textit{The integration of highlighting and summarization functions distinguishes it from other social media platforms.
It is more convenient for me to organize what I have learned in the same interface of the community without the need for additional software}''.
\revision{
Five participants mentioned that when they couldn't find relevant information in the community, they used the \question{} panel for additional support, making the information-seeking process more engaging.
P18 (M, 29) shared, ``\textit{When I couldn’t find answers in the comments, I used the panel to ask specific questions. It helped me reflect more deeply on my problem. Without it, I might have overlooked the issue, but instead, it encouraged me to keep exploring}''.
}

\subsection{Perceptions towards \name{} (RQ3)}
In general, participants show significantly more positive responses to the usefulness of \name{} ($Mdn = 6.00$) than baseline ($Mdn = 3.00$); $Z = -3.936, p < 0.001$. 
Participants also find that \name{} ($Mdn = 6.00$) is significantly easier to use than the baseline interface ($Mdn = 3.50$); $Z = -2.157, p = 0.031$. 
Overall, they have a significantly stronger intention to use \name{} ($Mdn = 6.00$) rather than the baseline interface ($Mdn = 5.00$) when they want to seek social support to help 
\revision{cope with their mental health concerns}
by browsing the content in OMHC; $Z = -3.311, p < 0.001$.
Participants generally agree that each of \name{}'s features in the \vis{} panel, the \note{} panel, and the \question{} panel is useful in their support-seeking tasks in the OMHC, as shown in \autoref{tab:feature}.
We summarize participants' qualitative responses to each 
\proofreading{panel}
as below. 

\subsubsection{It is 
\revision{simplified}
to find desired posts and comments via the \vis{} panel}
As reported in \autoref{sec:rq1_result} (RQ1) and \autoref{sec:rq2_result} (RQ2), the \vis{} panel allows participants to explore the posts based on topics in a zoomable pattern and filter posts and comments based on social support, which 
\revision{simplifies}
and makes it engaging
to articulate the mental health issues and find needed information in the OMHC. 
This benefit is explicitly appreciated by seventeen participants. 
Six of them especially favor the support filtering view that helps them find needed social support.
P15 (M, 22) said,  ``\textit{Many people when experiencing mental health issues may want to seek advice online, but for me, it would be more important to get emotional support as the advice online is often not practical. I therefore select the posts that seek for a large amount of emotional support and further filter comments that provide a large amount of emotional support as well.}''. \label{sec: emotional}

\subsubsection{It is convenient to manage the useful content via the \note{} panel}
As reported in \autoref{sec:rq1_result} (RQ1) and \autoref{sec:rq2_result} (RQ2), participants indicate that the \note{} panel enables them to highlight any content of their interest and get an automatic summary on the highlighted content, which reduces their cognitive workload and improves their support-seeking outcome. 
The benefits of the \note{} panel are indicated by seventeen participants. 
As another example, P13 (F, 22) stated, 
``\textit{I can categorize useful information using different colors, which provides clarity. 
Additionally, the tool can automatically summarize my collection and create a mind map, making it easy to review the useful information and form a take-away}''. 
Five participants also like the feature of navigating to previously highlighted content. 
P10 (M, 23) commented, 
``\textit{I have a habit of taking notes when I read something useful online, but when I review my notes, I often struggle to recall what they were based on. 
This tool enables me to easily locate the posts and comments from which my note is sourced, which helps me to review the content}''.

\subsubsection{It is inspiring to interact with the user-generated content and machine-generated content via the \question{} panel}
As reported in \autoref{sec:rq1_result} (RQ1), the \question{} panel is deemed useful to help participants make sense of the suggestions and improve their support-seeking outcome. 
In fact, five participants agree that the \question{} panel is one of the most helpful panels in \name{}.
\shiwei{For example, P1 (F, 22) stated,
``\textit{The answer helped me understand the confusing content. It makes it easier for viewers like us who are anxious to get support}''. 
}
P2 (F, 22) explained \zhenhui{the helpfulness of the \question{} panel}.
``\textit{The recommended questions really helped me analyze my mental health issue from multiple perspectives and inspired me to ask more questions. 
In this interactive process, I learned a lot about how to mitigate my mental health issue}''.
P8 (F, 23) said, ``\textit{Sometimes I struggle to articulate my questions about my confusions. The suggested questions happen to be what I want to ask}''. 
P19 (M, 23) also stated,
``\textit{The hierarchy generated questions and answers record my process of digging into one specific point gained from the content in the community. 
While exploratory with the suggestions, I may think of some concerns about the previous answer and return to where it is to add a new question.
After several rounds of interaction, I have a clearer mind on the current mental health issue and how to address it}''.

\section{Discussion} \label{sec:discussion}
In this work, we propose \name{}, an interactive visual tool to help viewers seek social support from the posts and comments in online mental health communities (OMHCs).
\revision{
Inspired by previous visual analysis systems for online community administrators, we designed a \vis{} panel to help users with mental health concerns navigate their needed social support with a simplified pattern.
Recognizing the capability of LLMs to generate instant, high-quality responses and drawing insight from prior research \cite{sensemaking24Gero} in LLM sensemaking, we explored how AI-generated content can support users in making sense of user-generated content in OMHCs by tailoring prompts to fit the specific needs and context of OMHCs.
}
Our within-subjects study with 20 participants showcases that \name{} 
significantly improves users' support-seeking outcome and engagement in the support-seeking task, compared with a baseline OMHC interface. 
Participants mainly attribute the improvements to \name{}'s \vis{} panel, which facilitates them to find desired posts and comments efficiently and helps them slow down and articulate their thoughts during the zoom-in and zoom-out process.
This finding aligns with previous work that providing a holistic picture of the search results and filtering features appropriate to the current scenario can aid users in easily navigating to the desired information \cite{shi2022medchemlens, peltonen2017topic, tu2022phrasemap}.
\revision{This highlights the need to provide viewers with mental health concerns a more targeted and less overwhelming method for searching, rather than requiring them to click through and read posts one by one from an exhaustive search list.}
Participants also favor the \note{} panel which makes it convenient and reduces the mental effort required to organize social support information to 
\revision{cope with the mental health concerns.}
This finding extends previous tools that use highlighting for information management \cite{10.1145/3555555, liu2023coargue, 10.1145/3586183.3606816, peng2020exploring} by providing evidence on its benefit in using highlighting to inform color-code organization and automatic summary \cite{summit24Gilmer}.
\shiwei{
In addition, the \question{} panel provides viewers with direct answers to their confusion and recommends possible questions to inspire their interaction and facilitate the \textit{digest} step.
Both the \note{} and the \question{} panel demonstrate the potential of machine-generated content to assist viewers in making sense of online user-generated content, \eg posts and comments in OMHCs, thereby alleviating their mental health 
\revision{concerns}.
\revision{These two panels help participants process their emotions and mental health concerns, encouraging deeper reflection and engagement by allowing them to summarize and pose questions.}
This finding aligns with previous works that providing some initiative action (\eg recommend question, automatic summary) can improve engagement and increase the learning outcome \cite{readerquizzer2023Liam, peng2023designquizzer, 10.1145/3586183.3606816, 10.1145/3586182.3615785}.
}
\zhenhui{
\subsection{Generality of \name{}}
\name{} can be easily generalized to other similar OMHCs (\eg r/Depression, 7cups) as the needs and classifiers for informational and emotional support are applicable to general OMHCs \cite{sharma2018mental}.
\revision{\name{} has its uniqueness in OMHCs.
Unlike general forums like Stack Overflow, which primarily address specific problems (\eg ``How to zoom around an SVG''), mental health forums provide both informational and emotional support.
The \vis{} panel enables users to identify the type of social support relevant to their needs.
In OMHCs, suggestions are often fragmented and presented as diverse, parallel perspectives rather than forming a cohesive solution, as noted by S05 (\autoref{sec: finding}).
The \note{} panel supports viewers in organizing the selected insights, helping transform fragmented information into more actionable, high-level suggestions.
The \question{} panel incorporates tailored prompts specifically designed for mental health contexts, enabling it to provide relevant responses and suggest questions that align more closely with the needs of OMHC users.}
While our proposed \name{} targets viewers and OMHCs in this paper, its \vis{}, \note{}, and \question{} panel can also be easily generalized to other types of online communities
(\eg design communities \cite{peng2023designquizzer}, question-answering communities \cite{10.1145/3555555}) \revision{ with certain adaptations}.  
In these online communities, users, whether they directly participate in the communication or not, can find and digest helpful information (\eg design knowledge, critical opinions) in the posts and comments. 
To implement a \vis{} panel in those communities, developers could conduct topic modeling on the posts and identify the most similar post pairs as we did. 
The support filtering view can be adapted to other filtering criteria by training classifiers based on what users care about the posts and comments, \eg the types of design examples shared in the post and the type of UI components mentioned in the comments \cite{peng2023designquizzer}. 
\revision{
The \note{} panel could be adapted to other communities by adjusting the summary format to align with the specific needs of the users in the community.
For instance, in technical forums (\eg Stack Overflow), it could shift from capturing fragmented insights to focusing on summarizing key solution steps in a clear, step-by-step format, aligning with the structured problem-solving processes common in these communities.
The \question{} panel could be tailored to different online communities by adjusting prompts to generate AI responses that are specifically aligned with the needs and focus of the community, \eg emphasizing design principles for design communities or offering technical solutions for question-answering communities.
}
\revision{
Beyond online communities, the unique design of these panels, originally developed to facilitate sensemaking in OMHCs, could also be extended to support other cognitively demanding tasks. 
As suggested by two participants, these panels could assist in tasks that require deep reflection and critical thinking, such as reading scientific papers \cite{peng2022crebot, 10.1145/3586183.3606816}.
}
}
\subsection{Design Considerations}
Based on our findings, we outline three design considerations for facilitating support-seeking in online mental health communities, or more broadly, general information-seeking tasks in \zhenhui{online} communities. 
\zhenhui{
\subsubsection{Support hierarchical visual interaction with the textual content} 

All of our three panels in \name{} support users to interact with the textual content in a hierarchical visual manner, which is found helpful in our study (\autoref{tab:feature}).
For example, the \vis{} panel allows users to zoom in and zoom out the visual circles to explore the textual posts. 
By exploring posts in different circle layers, users can view the returned posts of searched keywords from a clear and comprehensive perspective and clarify their thoughts on their desired posts and comments in OMHCs.
Besides, our \note{} panel groups the highlighted content based on the users' assigned colors and generates a mind map that summarizes the highlighted content in a hierarchical visual way. 
In addition, 
the \question{} panel assists viewers in interacting with the questions and answers in a hierarchy layout, facilitating a flexible interaction with user-generated content.
The finding aligns with the previous work that displaying information in a non-linear manner can reduce cognitive needs for comprehension \cite{wheeldon2011picture, larkin1987diagram, wheeldon2011picture}.
Enabling graphical and hierarchy layout of the textual content facilitates information exploration, organization and sensemaking \cite{suh2023sensecape, jiang2023graphologue}.
Thus, we recommend that future tools aiming to facilitate non-linear information-seeking tasks (\eg support-seeking in online communities) could support hierarchical visual interaction with features similar to \name{} (\eg non-linear layout, color-code organization) with textual content.
}

\subsubsection{Enhance sense of emotional support with user-generated and machine-generated content}
While our study shows that \name{} helps viewers to get significantly more key points about informational support from the OMHC compared to the baseline condition without \name{} (\autoref{tab:userStudyResults}), we \proofreading{found} that viewers seldom \proofreading{recorded} down the points that provide emotional support (\eg encouragement, comfort). 
One possible reason could be that the \note{} panel may shift participants' attention to seeking informational support in the community, as suggested by P17 (F, 22),
``\textit{Highlighting useful information while reading makes me focus more on some practice that helps alleviate the mental health issue}''.
\revision{
Another possible reason could be that participants tended to seek informational support rather than emotional support. 
}
Nevertheless, the comments, one type of user-generated content, in OMHCs commonly convey emotional support \cite{peng2020exploring, sharma2018mental}, which is also important to help alleviate viewers' mental health concerns \revision{as suggested by P15 (M, 22) (\autoref{sec: emotional})}.
\shiwei{
Previous works suggest that people are more satisfied when their received support matches the needed support \cite{vlahovic2014support, wolff2013physical}. 
}
To help viewers better sense the emotional support from the explored comments, future tools could automatically highlight the sentences that convey emotional support, which would catch viewers' eyes when they read the comments. 
Furthermore, \name{} currently uses LLMs to generate questions and answers in a technical format based on users' selected content. 
In the future, tools like \name{} could prompt LLMs to also act as emotional supporters \cite{wang2021cass} if viewers want to engage with it during the support-seeking process in online mental health communities.

\zhenhui{



\subsubsection{Support viewers with a smooth information-seeking process to reduce mental efforts}
In many learning and data exploration tasks, users often are uncertain about the boundaries of what text may be relevant and useful later \cite{chang2016supporting}. 
During the information-seeking process, users often lose track \cite{liu2022wigglite}, because making decisions or the trade-off between information from excessive online search results can quickly exceed the limits of working memory \cite{hinckley2012informal, liu2019unakite}.
\name{} provides features that help viewers easily track their performance during the support-seeking process, which reduces their cognitive load, as revealed by our RQ2 findings. 
For example, our \vis{} panel allows viewers to interact with posts in a zoomable pattern, thereby facilitating their ability to observe their exploration process.
The \note{} panel helps viewers navigate to previously highlighted content, thereby facilitating a more nuanced comprehension of the surrounding context and recall of one's own responses upon initial marking. 
In addition, the automatically generating summary about the suggestions accelerated users’ information processing.
It aligns with previous work \cite{summit24Gilmer} that providing automatic summarization can support users to acquire information from lengthy discussions with lower manual effort required.
The recommended questions and answers provided by LLMs assist viewers in being more engaged with the user-generated content which facilitates the digesting step in the support-seeking process.
It aligns with insights from previous work \cite{peng2023designquizzer} that structured summarization can ease the reading workload and quiz-like interaction that encourages active thinking.
Therefore, to smooth viewers' mental effort in the information-seeking process in online communities, future tools could consider the design and usage of tracking features (\eg navigating to the previous marking, visualizing the exploration process) and automatic features (\eg automatic summarizing, recommending questions). 
}

\subsection{Limitation and Future Work}
\label{sec: limitation}
\subsubsection{Limitations on clustering algorithm}
To support real-time clustering of the returned posts of the searched keywords,
we utilized LDA and preset the number (4 in our case) of clusters. 
However, this may result in some less meaningful clusters, which had been noted by three participants that they have difficulty distinguishing between multiple categories of posts quickly.
Future work can explore other clustering algorithms, \eg the dynamic clustering algorithm \cite{knittel2021real}, which can automatically determine the number of suitable clusters while maintaining an acceptable response time of the system. 

\subsubsection{Limitations on user study}
First, although we had counterbalanced the task-system assignment, we could not avoid the novelty effect that may bias people's perceptions of the usefulness of our introduced new tools. 
\zhenhui{
Second, although our participants are university students, a representative user group who commonly experiences mentally challenging problems \cite{rooksby2019student, macaskill2013mental, bagroy2017social}, they can not represent all user groups that would like to seek social support in OMHCs. 
Future work needs to test \name{} with more diverse user groups in a long-term user study.
Third, although r/Anxiety is one of the representative OMHCs, there is a need to evaluate \name{} in other OMHCs in order to validate the generalizability of our findings.
Fourth, although our user study has demonstrated the usefulness of \name{} as a whole, it does not separately evaluate the usefulness of each panel in viewers' support-seeking exploration process. 
To systematically distinguish the effects of each panel in \name{}, future work needs to conduct an ablation study, \ie with baselines that deduct each of the 3 panels (\autoref{fig:all}).
Fifth, we do not involve the crowd-contributed tags from the community members in both OMHC interfaces. While this design choice can help us focus on the impact of \name{}'s unique features on support-seeking practice, it misses an opportunity to give a bottom-up and human-annotated way for users to navigate the community. 
Future work can consider embedding \name{} into a public OMHC to examine how its features can complement existing features of the OMHC in viewers' support-seeking practice.
}

\revision{
\subsubsection{Limitations on AI-generated content used in online communities}
In our user study, we observed that participants tended to use the \question{} panel when specific details were unavailable in the community.
While this facilitated timely feedback, the AI-generated content did not incorporate community data, potentially leading to user mistrust or harmful outcomes due to unverified information.
To address issues like hallucinations \cite{zhang2023siren}, future research could focus on integrating AI-generated content (AIGC) with user-generated content (UGC) to provide more accurate and contextually relevant responses.
In cases where relevant UGC is unavailable, the system could rely on AI-generated content while clearly indicating the absence of community data.
This approach would help users understand that the information may not reflect actual user experience and enable them to make informed decisions.
}


\section{Conclusion}
In this paper, we propose \name{}, an interactive visual tool to help viewers seek social support in online mental health communities (OMHCs). 
\name{} provides a \vis{} panel that helps users visually filter and explore helpful posts and comments, a \note{} panel that enables users to highlight and summarize any content of their interests, and a \question{} panel that assists users in making sense of the comments with a large language model. 
Our within-subjects study with 20 participants demonstrates that compared to the baseline condition without \name{}, \name{} can significantly improve the amount of received information support from an OMHC and offer users a significantly more engaging support-seeking experience. 
We provide design considerations for supporting online information-seeking and information-management tasks with visualization techniques and large language models.
\bibliographystyle{ACM-Reference-Format}

\appendix
\section{Prompts used for Note-taking panel and Questioning panel}
\label{sec:appendix}
\subsection{Summary}
\subsubsection{Prompt}
Please summarize the suggestions: \textbf{\{suggestions\}} given and output the results organized with "subtitle" and "content" that is corresponding to the subtitle, format like:``subtitle: Patience with Diagnosis; content: The patience with Diagnosis'', and return a title that describes all the content.
\subsubsection{Example Inputs}
Putting on some really nice music (like lofi) really helps too, focus on your body instead of your thoughts. Really push yourself to what you know you can safely handle even if you don't think you have the energy. Adding ambient music helps too. Drink some Chamomille tea. Drink some Chamomille tea. Taking a melatonin supplement an hour or two before bed helps me sometimes.

\subsubsection{Example Response}
\textbf{Title}: Strategies for Relaxation and Better Sleep.

\textbf{Subtitle}: Musical Environment for Focus and Relaxation.
\textbf{Content}: Incorporating calming music, such as lofi, creates a soothing atmosphere that can help shift focus from intrusive thoughts to physical sensations. Adding ambient sounds further enhances this environment, promoting relaxation.
     
\textbf{Subtitle}: Physical Activity Within Safe Limits.
\textbf{Content}: Engaging in physical activities, pushing oneself to the extent that is safely manageable, can help in expending energy which might otherwise contribute to restlessness or anxiety.

\textbf{Subtitle}: Herbal Remedies for Calmness.
\textbf{Content}: Consuming Chamomile tea, known for its calming properties, can be a natural way to ease the mind and prepare the body for rest.
     
\textbf{Subtitle}: Sleep Enhancement with Supplements.
\textbf{Content}: Taking a melatonin supplement approximately one to two hours before bedtime can assist in regulating sleep patterns, making it easier to fall asleep.

\subsection{Recommended Questions}
\subsubsection{Prompt}
As someone with mental health issues, please ask three questions about the \textbf{\{current statement\}} from three different perspectives: what, why and how to do.
\subsubsection{Example Inputs}
Try to focus on relaxing your muscles and additionally.
\subsubsection{Example Response}
\textbf{Question1}: From a psychological perspective, how does focusing on relaxing muscles contribute to alleviating symptoms of anxiety or stress?

\textbf{Question2}: From a physical therapy point of view, could you suggest specific muscle relaxation techniques that are safe for individuals with pre-existing conditions?

\textbf{Question3}: From a holistic health standpoint, what complementary practices might enhance the benefits of muscle relaxation for mental well-being?

\subsection{Answers}
\subsubsection{Prompt}
As someone with expertise in mental health, please provide a brief answer to the question.
\subsubsection{Example Inputs}
From a physical therapy point of view, what specific muscle relaxation exercises would you recommend for someone with chronic tension?
\subsubsection{Example Response}
From a physical therapy standpoint, I would recommend progressive muscle relaxation, where you systematically tense and then relax different muscle groups, and deep breathing exercises combined with gentle stretching to relieve chronic muscle tension. Additionally, incorporating mindfulness-based practices like yoga or tai chi can also help to reduce overall stress and tension in the muscles.
\end{document}